\documentclass[12pt, leqno]{article}
\usepackage[pdftex,colorlinks=true]{hyperref}
\usepackage{amsmath,setspace,multirow,lineno,xcolor}
\usepackage[font=small,labelfont=bf]{caption}
\usepackage[top=1.3in, bottom=1.3in, left=1.1in, right=1.1in]{geometry}

\usepackage[round]{natbib}
\usepackage{color,soul}
\definecolor{darkblue}{rgb}{0,0,.6}
\hypersetup{citecolor=darkblue,linkcolor=darkblue,urlcolor=darkblue}
\usepackage{comment}
\usepackage{orcidlink}
\usepackage[linesnumbered,ruled,vlined]{algorithm2e}

\DeclareCaptionStyle{italic}[justification=centering]{labelfont={bf},textfont={it},labelsep=colon}
\usepackage{graphicx,psfrag,epsf,textcomp,epstopdf, amsthm, pdflscape}
\usepackage{enumerate, dsfont, alltt, verbatim}
\usepackage{natbib}
\usepackage{url}
\usepackage{booktabs, subfig, bm, paralist,mathpazo,tikz,todonotes,longtable,microtype}

\definecolor{americanrose}{rgb}{1.0, 0.01, 0.24}

\newcommand{\blind}{0}
\graphicspath{{plots/}}

\newcommand{\X}{\mathcal{X}}
\newcommand{\Y}{\mathcal{Y}}

\newcommand{\Rlogo}{\protect\includegraphics[height=1.8ex,keepaspectratio]{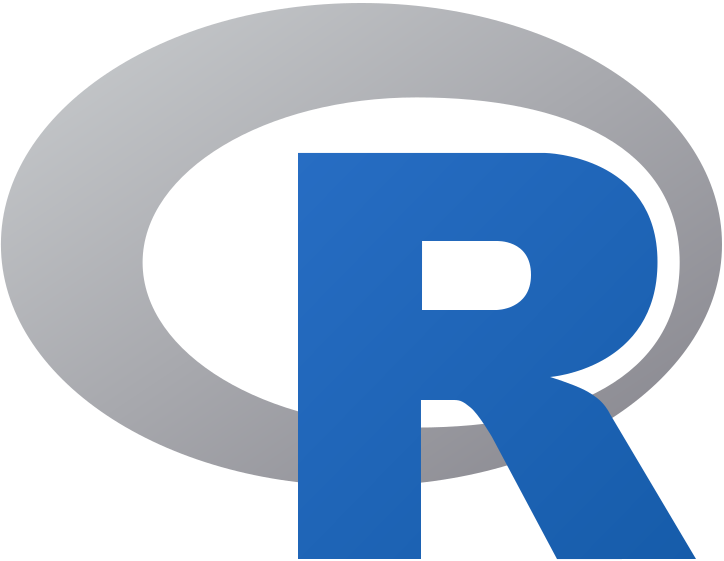}}

\newsavebox\CBox

 \newtheorem{@definition}{\sc Definition}[section]

\renewcommand\X{\mathcal{X}}

\date{}

\begin{document}

\def\spacingset#1{\renewcommand{\baselinestretch}{#1}\small\normalsize} \spacingset{1}

\if0\blind
{
\title{\bf \mbox{Penalized function-on-function linear quantile regression}}}
\author{
\normalsize Ufuk Beyaztas\footnote{Corresponding address: Department of Statistics, Marmara University, 34722, Kadikoy-Istanbul, Turkiye; Email: ufuk.beyaztas@marmara.edu.tr} \orcidlink{0000-0002-5208-4950} 
\\
\normalsize Department of Statistics \\
\normalsize Marmara University \\
\\
\normalsize Han Lin Shang \orcidlink{0000-0003-1769-6430} 
\\
\normalsize     Department of Actuarial Studies and Business Analytics \\
\normalsize     Macquarie University \\
\\
\normalsize Semanur Saricam \orcidlink{0000-0002-0580-4812}
\\
\normalsize Department of Statistics \\
\normalsize Marmara University
}
\maketitle
\fi

\if1\blind
{
\title{\bf Penalized function-on-function linear quantile regression}
} \fi

\maketitle

\begin{abstract}
We introduce a novel function-on-function linear quantile regression model to characterize the entire conditional distribution of a functional response for a given functional predictor. Tensor cubic $B$-splines expansion is used to represent the regression parameter functions, where a derivative-free optimization algorithm is used to obtain the estimates. Quadratic roughness penalties are applied to the coefficients to control the smoothness of the estimates. The optimal degree of smoothness depends on the quantile of interest. An automatic grid-search algorithm based on the Bayesian information criterion is used to estimate the optimum values of the smoothing parameters. Via a series of Monte-Carlo experiments and an empirical data analysis using Mary River flow data, we evaluate the estimation and predictive performance of the proposed method, and the results are compared favorably with several existing methods.
\end{abstract}


\newpage
\spacingset{1.55} 

\section{Introduction} \label{sec:intro}

Classical regression analysis models focus on the center of the conditional distribution of response. When the response's non-central locations (i.e., tails) are of interest, the quantile regression proposed by \cite{Koenker1978} is a powerful tool. Quantile regression has several noteworthy advantages over conditional mean regression: 
\begin{inparaenum}
\item[1)] By characterizing the entire conditional distribution of the response, quantile regression provides a complete picture of the connectivity between the response and predictors, which cannot be reflected by conditional mean regression.  
\item[2)] Quantile regression pertains to a robust model family and, thus, it is more robust to outliers in the response variable than conditional mean regression. 
\item[3)] Compared with conditional mean regression, quantile regression provides more efficient inference when the error terms follow a non-symmetric heavy-tailed distribution or in conditional heteroskedasticity. 
\end{inparaenum}
Therefore, quantile regression has become a popular tool, and it has successfully been applied in many real-life scenarios \citep[see, e.g.,][]{Eilers2012, Briollais2014, Magzamen2015, Lehn2018, Abbas2019, Lara2019, Vasseur2021}.

The quantile regression discussed above is best suited for analyzing the data in a discrete data matrix. In recent years, technological breakthroughs in data collection and storage have given rise to observations that can be recorded over a continuous grid, such as age, space, time and wavelength. These stochastic processes, called functional data, bring many challenges and opportunities because of their infinite-dimensional nature. Classical statistical models are not practical for analyzing functional data. Therefore, developing valid approaches to analyzing functional data has received considerable attention. The work of \cite{ramsay2002}, \cite{ferraty2006}, \cite{ramsay2006}, \cite{horvath2012}, \cite{cuevas2014}, \cite{Hsing2015}, and \cite{Kokoszka2017} provide a general overview, theoretical developments, and case studies in functional data analysis.

In the functional data analysis literature, particular attention has been paid to the functional linear model, in which at least one response or predictor variable consists of infinite-dimensional random curves. Several functional linear quantile regression models have been proposed by extending quantile regression setting into functional data. For scalar-on-function regression models, where the response is scalar-valued, and the predictor involves random curves, for example, \cite{cardot2005, ferraty2005, cardot2007, chenmuller2012, Kato2012, Tang2014, Yu2016, yao2017, Ma2019, Sang2020, Chaouch2020, Zhang2021, Li2022, Zhu2022, Zhou2022}. For function-on-scalar regression models, where the response consists of random curves and the predictor consists of scalar values, for example, \cite{Kim2007, Wang2009, yang2020, Liu2020}. 

Compared with scalar-on-function and function-on-scalar regression models, quantile regression has been studied much less for the function-on-function regression models, where both the response and predictor involve random curves, with two exceptions as follows. \cite{BSA2022} propose a function-on-function linear quantile regression model by adapting the partial quantile regression approach of \cite{Dodge} into the functional partial least squares regression setting. \cite{BSMMA} propose a functional linear quantile regression model by transforming all the functional objects in the model into a finite-dimensional space via the functional principal component analysis paradigm. 

While these two approaches are intuitive and consistent \citep{Hall2007}, they have some drawbacks. For both methods, estimating the number of components is difficult. The shape and interpretation of the functional parameter can change when one includes one or two additional partial least squares and principal components \citep{Crainiceanu2009, ivanescu2015}. In both methods, the basis dimension for the functional predictor induces the smoothness of the functional parameter, which can result in significant under-smoothing if the functional parameter is considerably smoother than the higher-order partial least squares and principal component scores \citep[see, e.g.,][]{ivanescu2015}.

As demonstrated by \cite{BSA2022} and \cite{BSMMA}, quantile regression provides more robust estimates than mean regression when outliers are present in the functional response. However, as discussed in Section~\ref{sec:mc} in detail, our numerical results demonstrate that the existing quantile regression models may not be adequately robust to outliers when the underlying regression model has a wavy parameter function. This phenomenon leads to a more fluctuated functional response. An explanation is that unpenalized quantile regression estimators result in biased parameter estimates when the data have both a wavy regression parameter function and outliers. Thus, the robustness degree of the existing estimators is limited in such cases.

In this paper, we propose a novel penalized function-on-function linear quantile regression model. In our model, similar to the penalized function-on-function regression proposed by \cite{ivanescu2015}, the regression parameter functions are represented by tensor cubic $B$-splines expansion. Compared with penalized linear models, deriving the form of the penalized quantile regression model is very challenging, as the quantile loss function is not differentiable everywhere. To tackle this problem, we consider a smoothed loss function that dominates the original quantile loss function. 

In the model, the quadratic roughness penalties are applied to the expansion coefficients to control the smoothness of the estimates. To obtain the estimates of the regression parameter functions, we consider a derivative-free optimization algorithm that is computationally efficient and converges quickly to optimum values. In the algorithm, the regression parameter function estimate is defined as the optimizer of a form of smoothed and penalized loss function. Compared with unpenalized estimators, which generally result in either under-smoothed or over-smoothed estimates of the regression coefficient functions, our estimator enforces a certain level of smoothness on the estimates to avoid overfitting. 

Our estimators increase the robustness of the existing unpenalized estimators when the underlying regression model has a wavy parameter function; i.e., compared with existing unpenalized estimators, our proposed estimator yields more robust inference when the regression parameter function is wavy and the data include outliers (refer to the numerical results in Section~\ref{sec:mc}). The optimal degree of smoothness depends on the quantile of interest. To determine the optimal smoothing parameter, we use an automatic grid-search algorithm based on the Bayesian information criterion (BIC). In the case of a scalar response, quantiles are uniquely defined. However, when the response is function-valued, unique quantiles are no longer guaranteed. One straightforward strategy for applying quantile regression to a functional response involves fitting independent quantile regressions for each continuum of the functional response. While this approach yields unbiased estimates, it is inefficient as it fails to leverage information from nearby continua, a common practice in functional data modeling methodologies \citep[see, e.g.,][]{yang2020, Liu2020}. Our proposed method enhances efficiency by incorporating B-spline basis functions and applying penalization to induce smoothness and regularization in the functional coefficient, allowing for the borrowing of strength across continuum of the functional response.

The rest of this paper is structured as follows. Section~\ref{sec:2} introduces the function-on-function linear quantile regression model. The procedure used to estimate the parameters of the model is presented in Section~\ref{sec:3}. A series of Monte-Carlo experiments and an empirical data analysis are performed to evaluate the finite-sample performance of the method. The results are tabulated in Sections~\ref{sec:mc} and~\ref{sec:data}. Section~\ref{sec:conc} concludes the paper, along with some ideas on how the methodology can be extended.

\section{Model, notations, and nomenclature}\label{sec:2}

Let $\Y$ and $\X$ denote square-integrable random processes defined on the compact subsets of $\mathbb{R}$, that is $\mathcal{I}_y \subseteq \mathbb{R}$ and $\mathcal{I}_x \subseteq \mathbb{R}$, respectively. Throughout this paper, both $\Y$ and $\X$ are assumed to be the elements of $\mathcal{L}_2$ Hilbert space $\mathcal{H}$. In what follows, we assume that we observe $n$ independent copies of $(\Y, \X)$; $\lbrace \Y_i(t), \X_i(s):~ i = 1, \ldots, n \rbrace $ where $ t \in \mathcal{I}_y$ and $s \in \mathcal{I}_x$. For a given quantile level $\tau \in (0,1)$, let $Q_{\tau}[\Y_i | \X_i ]$ denote the $\tau\textsuperscript{th}$ quantile of $\Y$ conditional on the entire trajectory $\X_i$. Then, the function-on-function linear quantile regression model considered in this paper is of the form
\begin{equation}\label{eq:fflqrm}
Q_{\tau}[\Y_i | \X_i] = \alpha_{\tau}(t) + \int_{\mathcal{I}_x} \X_i(s) \beta_{\tau}(t,s) ds,
\end{equation}
where $\alpha_{\tau}(t): \mathcal{I}_y \rightarrow \mathbb{R}$ is the intercept function and $\beta_{\tau}(t,s): \mathcal{I}_y \times \mathcal{I}_x \rightarrow \mathbb{R}$ is a two-dimensional regression coefficient function that evaluates the effect of $\X_i$ on the $\tau\textsuperscript{th}$ quantile of $\Y$.

The regularized estimate of $[\alpha_{\tau}(t), \beta_{\tau}(t,s)]$ in~\eqref{eq:fflqrm} can be obtained by minimizing the following objective function:
\begin{equation}\label{eq:obj1}
\underset{\begin{subarray}{c}
  (\alpha_{\tau}, \beta_{\tau}) \in \mathcal{H} 
  \end{subarray}}{\arg\min}~ \sum_{i=1}^n \rho_{\tau} \left[ \Y_i(t) - \alpha_{\tau}(t) - \int_{\mathcal{I}_x} \X_i (s) \beta_{\tau}(t,s) ds \right] + \frac{\lambda_1}{2} \mathcal{J}_1(\alpha_{\tau}) + \frac{\lambda_2}{2} \mathcal{J}_2(\beta_{\tau}),
\end{equation}
where $\rho_{\tau}(u) = [\tau - \mathbb{1}(u < 0)]$ with indicator function $\mathbb{1}(\cdot)$ is the usual check loss function introduced by \cite{Koenker1978}, $\mathcal{J}_1$ and $\mathcal{J}_2$ are the roughness penalties on $\alpha_{\tau} \in \mathcal{H}$ and $\beta_{\tau} \in \mathcal{H}^2$, respectively, and $\lambda_1, \lambda_2 > 0$ are the smoothing parameters that control the amount of shrinkage in~$\alpha_{\tau}$ and~$\beta_{\tau}$, respectively.

\section{Parameter estimation}\label{sec:3}

Similar to \cite{ivanescu2015} and \cite{Cai2021}, we assume that the $i\textsuperscript{th}$ elements of $\Y(t)$ and $\X(s)$ are densely observed such that $\Y_i(t) = \Y_i(t_{ij})$ and $\X_i(s) = \X_i(s_{ir})$ for $j = 1, \ldots, M_i$ and $r = 1, \ldots, G_i$, where $M_i$ and $G_i$ denote the numbers of observations for curves $\Y_i(t)$ and $\X_i(s)$, respectively. Without loss of generality, we consider the case where $t_{ij} = t_j$ ($j = 1, \ldots, M$) and $s_{ir} = s_r$ ($r = 1, \ldots, G$). We assume that the intercept function is represented by the linear combinations of $B$-spline basis function expansion with truncation constant $K_0$ as follows:
\begin{equation}\label{eq:bspa}
\alpha_{\tau}(t) \approx \sum_{k=1}^{K_0} a_k(\tau) \phi_k(t),
\end{equation}
where, for $k = 1, \ldots, K_0$, $\phi_k(t) \in \mathcal{I}_y$ and $a_k(\tau)$ are the $B$-spline basis functions and the corresponding basis expansion coefficients, respectively. It is also assumed that the two-dimensional regression coefficient function is represented by tensor product $B$-spline basis function with truncation constants $K_y$ and $K_x$ as follows:
\begin{equation}\label{eq:bspb}
\beta_{\tau}(t,s) \approx \sum_{l=1}^{K_y} \sum_{p=1}^{K_x} b_{lp}(\tau) \psi_l(t) \vartheta_p(s),
\end{equation}
where, for $l = 1, \ldots, K_y$ and $p = 1, \ldots, K_x$, $\psi_l(t) \in \mathcal{I}_y$ and $\vartheta_p(s) \in \mathcal{I}_x$ are the $B$-spline basis functions and $b_{lp}(\tau)$ is the basis expansion coefficient. Let $\Delta_r$ denote the length of $r$\textsuperscript{th} interval in $\mathcal{I}_x$, that is $\Delta_r = s_{r+1} - s_r$. Then, we consider the numerical integration to approximate $\int_{\mathcal{I}_x} \X_i(s) \beta_{\tau}(s,t) ds$ in~\eqref{eq:fflqrm} as follows:
\begin{align}
\int_{\mathcal{I}_x} \X_i(s) \beta_{\tau}(t,s) ds & \approx \sum_{r=1}^{G-1} \Delta_r \beta_{\tau}(t, s_r) \X_i(s_r) \nonumber \\
& = \sum_{r=1}^{G-1} \Delta_r \sum_{l=1}^{K_y} \sum_{p=1}^{K_x} b_{lp}(\tau) \psi_l(t) \vartheta_p(s_r) \X_i(s_r) \nonumber \\
& = \sum_{l=1}^{K_y} \sum_{p=1}^{K_x} b_{lp}(\tau) \psi_l(t) \widetilde{\vartheta}_{p,i} \label{eq:numint},
\end{align}
where $\widetilde{\vartheta}_{p,i} = \sum_{r=1}^{G-1} \Delta_r \vartheta_p(s_r) \X_i(s_r)$. Substituting~\eqref{eq:bspa}--\eqref{eq:numint} in~\eqref{eq:fflqrm}, we have the following approximate function-on-function linear quantile regression model:
\begin{equation}
Q_{\tau}[\Y_i | \X_i] \approx \sum_{k=1}^{K_0} a_k(\tau) \phi_k(t) + \sum_{l=1}^{K_y} \sum_{p=1}^{K_x} b_{lp}(\tau) \psi_l(t) \widetilde{\vartheta}_{p,i}.
\end{equation}

We consider the penalty functionals $\mathcal{J}_1$ and $\mathcal{J}_2$ in~\eqref{eq:obj1}. Similar to \cite{ivanescu2015}, we employ quadratic penalties, which can be obtained using the vector of second derivatives of the $B$-spline basis functions. Let $\bm{a}(\tau) = [a_1(\tau), \ldots, a_{K_0}(\tau)]^\top$. Then, we approximate the penalty functional for the intercept function $\alpha_{\tau}(t)$, denoted by $\mathcal{J}_1(\alpha_{\tau})$, as follows:
\begin{equation}\label{eq:p1}
\widetilde{\mathcal{J}}_1(\alpha_{\tau}) = \int_{\mathcal{I}_y} [\alpha_{\tau}^{(2)}(t)]^2 dt = \bm{a}^\top(\tau) \bm{P}_{\alpha} \bm{a}(\tau),
\end{equation}
where $\alpha_{\tau}^{(2)}(t)$ is the second derivative of $\alpha_{\tau}(t)$ and $\bm{P}_{\alpha}$ is the $K_0 \times K_0$-dimensional penalty matrix of the $B$-spline basis functions, $(k k^{\prime})\textsuperscript{th}$ entries of which are equal to $P_{\alpha, k k^{\prime}} = \int_{\mathcal{I}_y} \phi_k^{(2)}(t) \phi_{k^{\prime}}^{(2)}(t) dt$ for $k, k^{\prime} = 1, \ldots, K_0$. Let $\bm{b}(\tau) = [b_{lp}(\tau)]_{lp}$ denote a $K_y \times K_x$-dimensional coefficient matrix. Then, we approximate the penalty functional for the two-dimensional regression coefficient function $\beta_{\tau}(t,s)$, $\mathcal{J}_2(\beta_{\tau})$, as follows:
\begin{align}\label{eq:p2}
\widetilde{\mathcal{J}}_2(\beta_{\tau}) &= \int_{\mathcal{I}_y} \int_{\mathcal{I}_x} \left[ \frac{\partial^2}{\partial t^2} \beta_{\tau}(t,s) \right]^2 ds dt + \int_{\mathcal{I}_y} \int_{\mathcal{I}_x} \left[ \frac{\partial^2}{\partial s^2} \beta_{\tau}(t,s) \right]^2 ds dt \notag\\
&= \bm{b}^\top(\tau) (\bm{\vartheta} \otimes \bm{P}_y + \bm{P}_x \otimes \bm{\psi}) \bm{b}(\tau),
\end{align}
where $\bm{\psi} = \int_{\mathcal{I}_y} \bm{\psi}(t) \bm{\psi}^\top(t) dt$ with $\bm{\psi}(t) = [\psi_1(t), \ldots, \psi_{K_y}(t)]^\top$, $\bm{\vartheta} = \int_{\mathcal{I}_s} \bm{\vartheta}(s) \bm{\vartheta}^\top(s) ds$ with $\bm{\vartheta}(s) = [\vartheta_1(s), \ldots, \vartheta_{K_x}(s)]^\top$, and $\bm{P}_y$ and $\bm{P}_x$ are the penalty matrices, $(l l^{\prime})\textsuperscript{th}$ and $(p p^{\prime})\textsuperscript{th}$ the entries of which are equal to $P_{y, l l^{\prime}} = \int_{\mathcal{I}_y} \psi_l^{(2)}(t) \psi_{l^{\prime}}^{(2)}(t) dt$ and $P_{x, p p^{\prime}} = \int_{\mathcal{I}_x} \vartheta_p^{(2)}(s) \vartheta_{p^{\prime}}^{(2)}(s) ds$ for $l, l^{\prime} = 1, \ldots, K_y$ and $p, p^{\prime} = 1, \ldots, K_x$, respectively.

Using the approximate penalty functionals in~\eqref{eq:p1} and~\eqref{eq:p2}, the estimates of $\bm{a}(\tau)$ and $\bm{b}(\tau)$ can be obtained by minimizing the following objective function:
\begin{equation}\label{eq:obj2}
\underset{\begin{subarray}{c}
  \bm{a}(\tau), \bm{b}(\tau)
  \end{subarray}}{\arg\min}~ \sum_{i=1}^n \sum_{j=1}^M \rho_{\tau} \left[ \Y_i(t_j) - \bm{\phi}^\top(t_j) \bm{a}(\tau) - (\bm{\widetilde{\vartheta}}_i^\top \otimes \bm{\psi}^\top (t_j)) \bm{b}(\tau) \right] + \frac{\lambda_1}{2} \widetilde{\mathcal{J}}_1(\alpha_{\tau}) + \frac{\lambda_2}{2} \widetilde{\mathcal{J}}_2(\beta_{\tau}),
\end{equation}
where $\bm{\phi}(t) = [\phi_1(t), \ldots, \phi_{K_0}(t)]^\top$ and $\bm{\widetilde{\vartheta}}_i = [\widetilde{\vartheta}_{1,i}, \ldots, \widetilde{\vartheta}_{K_x,i}]^\top$. Accordingly, the regularized estimates of $\alpha_{\tau}(t)$ and $\beta_{\tau}(t,s)$ can be computed as follows:
\begin{equation}\label{eq:est1}
\widehat{\alpha}_{\tau}(t) = \bm{\phi}^\top(t) \widehat{\bm{a}}(\tau), \qquad \widehat{\beta}_{\tau}(t,s) = (\bm{\vartheta}^\top (s) \otimes \bm{\psi}^\top (t)) \widehat{\bm{b}}(\tau),
\end{equation}
where $\widehat{\bm{a}}(\tau)$ and $\widehat{\bm{b}}(\tau)$ are the estimates of $\bm{a}(\tau)$ and $\bm{b}(\tau)$, respectively, obtained by minimizing~\eqref{eq:obj2}.

The estimates of $\bm{a}(\tau)$ and $\bm{b}(\tau)$ can be obtained using the penalized iteratively-reweighted least squares algorithm proposed by \cite{Nychka1995}. However, the quantile loss function used in this algorithm to compute $\widehat{\bm{a}}(\tau)$ and $\widehat{\bm{b}}(\tau)$ is not differentiable everywhere. To overcome this problem, \cite{Nychka1995} suggest using an approximate loss function differentiable everywhere. While this approach works well for the quantile levels around $\tau = 0.5$, it fails to converge for extreme quantile levels. Thus, we consider a gradient descent algorithm to compute $\widehat{\bm{a}}(\tau)$ and $\widehat{\bm{b}}(\tau)$. The gradient-based methods with the quantile loss function $\rho_{\tau}$ are not directly applicable to solve the optimization problem in~\eqref{eq:obj2} because $\rho_{\tau}$ is not differentiable everywhere. Thus, we consider the approximate check loss function proposed by \cite{Zheng2011} because of its ability to approximate the original quantile loss function quickly. 

The function-on-function linear quantile regression projected onto the finite-dimensional space, that is \eqref{eq:obj2} can be considered a quantile regression model for clustered (repeated) data. From this perspective, the estimates of $\bm{a}(\tau)$ and $\bm{b}(\tau)$ can also be obtained by the two-step estimation methodology proposed by \cite{BST+22}. This method first predicts the cluster-specific random effects using a linear quantile mixed model. Then, these predictions are used as offsets in the standard quantile regression model to estimate the regression parameters. While both the method of \cite{BST+22} and our proposed method can be used for estimating the regression coefficients in the finite-dimensional space, our experiences show that the former may not be appropriate for high-dimensional data. This is because the between columns of basis expansion coefficients become highly correlated, also known as multicollinearity, due to the nature of functional data, and the linear quantile mixed model used in the first stage of the method of \cite{BST+22} fails to provide estimates due to multicollinearity. 

In the function-on-function linear quantile regression model, we assume that the response and predictor are observed at fine grids and without measurement error. However, in most of the empirical applications, the data are observed with error, and, in these cases, a single penalty in the functional parameter estimation of the original model, as in our proposed method, may not be enough. In such a case, a pre-smoothing step of the sample curves of the functional predictor and response may be needed.

\subsection{Smoothed objective function}\label{sec:3.1}

The approximate check loss function introduced by \cite{Zheng2011} is of the form
\begin{equation}\label{appclf}
\widetilde{\rho}_{\tau, \gamma}(u) = \tau u + \gamma \ln(1 + e^{- \frac{u}{\gamma}}),
\end{equation}
where $\gamma$ is the tuning parameter. For any given $\gamma > 0$, it has been shown that $\widetilde{\rho}_{\tau, \gamma}(u)$ is a convex function and $\lim_{\gamma \rightarrow 0^+} \widetilde{\rho}_{\tau, \gamma}(u) = \rho_{\tau}(u)$ \citep[for more details, see Lemmas~1 and~2 in][]{Zheng2011}. Using $\widetilde{\rho}_{\tau, \gamma}(u)$, we estimate the parameters $\bm{a}(\tau)$ and $\bm{b}(\tau)$ in~\eqref{eq:obj2} as follows:
\begin{equation*}
\widehat{\bm{a}}(\tau),~ \widehat{\bm{b}}(\tau) = \underset{\begin{subarray}{c}
  \bm{a}(\tau), \bm{b}(\tau)
  \end{subarray}}{\arg\min}~ \Phi[\bm{a}(\tau), \bm{b}(\tau)],
\end{equation*}
where
\[
\Phi[\bm{a}(\tau), \bm{b}(\tau)] = \sum_{i=1}^n \sum_{j=1}^M \rho_{\tau} \left[ \Y_i(t_j) - \bm{\phi}^\top(t_j) \bm{a}(\tau) - (\bm{\widetilde{\vartheta}}_i^\top \otimes \bm{\psi}^\top (t_j)) \bm{b}(\tau) \right] + \frac{\lambda_1}{2} \widetilde{\mathcal{J}}_1(\alpha_{\tau}) + \frac{\lambda_2}{2} \widetilde{\mathcal{J}}_2(\beta_{\tau}).
\]
We define
\begin{equation*}
\widehat{\bm{a}}(\tau, \gamma),~ \widehat{\bm{b}}(\tau, \gamma) = \underset{\begin{subarray}{c}
  \bm{a}(\tau), \bm{b}(\tau)
  \end{subarray}}{\arg\min}~ \Phi_{\gamma}[\bm{a}(\tau), \bm{b}(\tau)],
\end{equation*}
where 
\begin{equation}
\Phi_{\gamma}[\bm{a}(\tau), \bm{b}(\tau)] = \sum_{i=1}^n \sum_{j=1}^M \widetilde{\rho}_{\tau, \gamma} \left[ \Y_i(t_j) - \bm{\phi}^\top(t_j) \bm{a}(\tau) - (\bm{\widetilde{\vartheta}}_i^\top \otimes \bm{\psi}^\top (t_j)) \bm{b}(\tau) \right] + \frac{\lambda_1}{2} \widetilde{\mathcal{J}}_1(\alpha_{\tau}) + \frac{\lambda_2}{2} \widetilde{\mathcal{J}}_2(\beta_{\tau}). \label{eq:sqr}
\end{equation}
The function $\Phi_{\gamma}[\bm{a}(\tau), \bm{b}(\tau)]$ represents the smoothed objective function, and the quantile regression model obtained by $\Phi_{\gamma}[\bm{a}(\tau), \bm{b}(\tau)]$ represents the smooth quantile regression model with the estimated parameters $\widehat{\bm{a}}(\tau, \gamma)$ and $\widehat{\bm{b}}(\tau, \gamma)$. From Theorem 1 of \cite{Zheng2011}, it is clear that $\Phi[\bm{a}(\tau), \bm{b}(\tau)]$ and $\Phi_{\gamma}[\bm{a}(\tau), \bm{b}(\tau)]$ are convex functions of $\widehat{\bm{a}}(\tau)$ and $\widehat{\bm{b}}(\tau)$, such that $[\widehat{\bm{a}}(\tau),~ \widehat{\bm{b}}(\tau)]$ and $[\widehat{\bm{a}}(\tau, \gamma),~ \widehat{\bm{b}}(\tau, \gamma)]$ exist and $[\widehat{\bm{a}}(\tau),~ \widehat{\bm{b}}(\tau)] \rightarrow [\widehat{\bm{a}}(\tau, \gamma),~ \widehat{\bm{b}}(\tau, \gamma)]$ as $\gamma \rightarrow 0^+$. The results given above indicate that small $\gamma$ values in the solution of the smooth quantile regression model approximate the original quantile regression model.

\subsection{Minimization of the smoothed objective function}\label{sec:3.2}

As in \cite{Zheng2011}, the smoothed objective function~\eqref{eq:sqr} can be minimized via gradient descent algorithms, such as the Newton and Broyden–Fletcher–Goldfarb–Shanno algorithms \citep[see, e.g.,][]{Walsh1975}. Such methods use the inverse of the Hessian matrix. Depending on the data structure, the Hessian matrix may be close to zero, and the information about the derivative of the objective may not be available (i.e., impractical to obtain or computationally infeasible). In such a case, these derivative-based methods may not produce a result \citep{Zheng2011}. For this reason, we consider the derivative-free algorithm ``Bounded Optimization BY Quadratic Approximation (BOBYQA)'' proposed by \cite{Powell2009} to minimize~\eqref{eq:sqr}. As is typical in general optimization techniques, BOBYQA, which utilizes a trust region method, constructs an approximation for the objective function by interpolation.

To simplify the notations, we consider the matrix representation of the smooth objective function. Let $\bm{\theta}(\tau) = [\bm{a}^\top(\tau), \bm{b}^\top(\tau)]^\top$, $\bm{\Pi} = [\bm{\Pi}_1, \ldots, \bm{\Pi}_n]^\top$ with $\bm{\Pi}_i = [\bm{\phi}^\top(t), \bm{\widetilde{\vartheta}}_i^\top \otimes \bm{\psi}^\top (t)]^\top$, and $\bm{P}(\lambda_1, \lambda_2)$ denote a block-diagonal matrix with elements $\lambda_1 \bm{P}_{\alpha}$ and $\lambda_2 (\bm{\vartheta} \otimes \bm{P}_y + \bm{P}_x \otimes \bm{\psi})$. Then, the smooth objective function can be represented as follows:
\begin{equation}\label{eq:rf}
\Phi_{\gamma}[\bm{\theta}(\tau)] = \sum_{i=1}^n \widetilde{\rho}_{\tau, \gamma} \left[ \Y_i(t) - \bm{\Pi}_i \bm{\theta}(\tau) \right] + \bm{P}(\lambda_1, \lambda_2) \bm{\theta}(\tau).
\end{equation}

For the minimization of~\eqref{eq:sqr}, the derivative-based trust-region methods usually start with constructing a local quadratic model at iteration $h$, $m^{[h]}(\bm{\eta}) \approx \Phi_{\gamma}[\bm{\theta}(\tau)^{[h]} + \bm{\eta}]$:
\begin{equation*}
m^{[h]}(\bm{\eta}) = \Phi_{\gamma}[\bm{\theta}^{[h]}(\tau)] + \nabla \Phi_{\gamma}^\top[\bm{\theta}^{[h]}(\tau)] \bm{\eta} + \frac{1}{2} \bm{\eta}^\top \bm{H}^{[h]} \bm{\eta},
\end{equation*}
where $\bm{\eta} \in \mathbb{R}^{v}$, $v$ is the number of variables in the model, and $\bm{H}^{[h]} \approx \nabla^2 \Phi_{\gamma}[\bm{\theta}^{[h]}(\tau)] \in \mathbb{R}^{v \times v}$ is the Hessian (symmetric) matrix of the objective or its approximation. Alternatively, the derivative-free trust-region methods construct the local model using interpolation instead of derivatives. Let $\bm{Z}^{[h]} \subset \mathbb{R}^v$ denote a set of interpolation points at iteration $h$. We present the general structure of the derivative-free trust-region methods in the minimization of~\eqref{eq:sqr} in Algorithm~\ref{alg:dfa}.

\begin{algorithm}[h]
\begin{small}
For $h = 0$, obtain the initial parameter estimates (i.e., $\bm{\theta}^{[0]}(\tau)$) using the penalized linear regression with given $\lambda_1$ and $\lambda_2$ values. \\
For $h = 1, 2, \ldots$ do \linebreak
\begin{scriptsize}
\textbf{2.1}
\end{scriptsize}
Use a polynomial approximation (usually a quadratic model) for the objective as follows:
\begin{equation*}
m^{[h]}(\bm{\eta}) = c^{[h]} + (\bm{g}^{[h]})^\top \bm{\eta} + \frac{1}{2} \bm{\eta}^\top \bm{H}^{[h]} \bm{\eta},
\end{equation*}
where $c^{[h]} \in \mathbb{R}$, $\bm{g}^{[h]} \in \mathbb{R}^v$, and $\bm{H}^{[h]} = (\bm{H}^{[h]})^\top$ satisfying the interpolation conditions $m^{[h]}[\bm{z} - \bm{\theta}^{[h]}(\tau)] = \Phi_{\gamma}[\bm{z}]$, $\forall~ \bm{z} \in \bm{Z}^{[h]}$ \citep[see, e.g.,][]{Powell2009}. \linebreak
\begin{scriptsize}
\textbf{2.2} 
\end{scriptsize}
Compute the step $\bm{\eta}$ by solving the trust region problem, i.e., $\bm{\eta}^{[h]}$ is obtained by solving
\begin{equation*}
\min_{\bm{\eta} \in \mathbb{R}^v} m^{[h]}(\bm{\eta}) ~~ \text{subject~to} ~~ \Vert \bm{\eta} \Vert \bigtriangleup^{[h]},
\end{equation*}
where $\bigtriangleup$ is the radius of the region where the model is considered as accurate and $\Vert \cdot \Vert$ is the Euclidean norm. \linebreak
\begin{scriptsize}
\textbf{2.3} 
\end{scriptsize}
Evaluate $\Phi_{\gamma}[\bm{\theta}(\tau)^{[h]} + \bm{\eta}^{[h]}]$ and compute $\ell^{[h]}$:
\begin{equation*}
\ell^{[h]} = \frac{\Phi_{\gamma}[\bm{\theta}(\tau)^{[h]}] - \Phi_{\gamma}[\bm{\theta}(\tau)^{[h]} + \bm{\eta}^{[h]}]}{m^{[h]}(\bm{0}) - m^{[h]}(\bm{\eta}^{[h]})}.
\end{equation*} \linebreak
\begin{scriptsize}
\textbf{2.4} 
\end{scriptsize}
If $\ell^{[h]}$ is sufficiently large, then set $\bm{\theta}(\tau)^{[h+1]} = \bm{\theta}(\tau)^{[h]} + \bm{\eta}^{[h]}$ and increase $\bigtriangleup^{[h]}$, otherwise, set $\bm{\theta}(\tau)^{[h+1]} = \bm{\theta}(\tau)^{[h]}$ and decrease $\bigtriangleup^{[h]}$. \linebreak
\begin{scriptsize}
\textbf{2.5} 
\end{scriptsize}
If $\ell^{[h]}$ is sufficiently large, then, update the $\bm{Z}^{[h]}$ by adding the point $\bm{\theta}(\tau)^{[h+1]}$ to $\bm{Z}^{[h]}$ and removing the point one far from $\bm{\theta}(\tau)^{[h+1]}$, otherwise, set $\bm{Z}^{[h+1]} = \bm{Z}^{[h]} \cup \lbrace \bm{\theta}(\tau)^{[h]} + \bm{\eta}^{[h]} \rbrace \setminus \lbrace \bm{z} \rbrace$ for some $\bm{z} \in \bm{Z}^{[h]}$ \citep[see, e.g.][]{Powell2009, Cartis2022}.
\end{small}
\caption{\small{Derivative-free trust-region optimization}}
\label{alg:dfa}
\end{algorithm}

In the BOBYQA method, the parameters $c$, $\bm{g}$, and $\bm{H}$ in Step 2.1 of Algorithm~\ref{alg:dfa} are updated at each iteration based on the change in the consecutive quadratic terms $\bm{H}^{[h]}$. In more detail, at each iteration, the parameters of the interpolating quadratic model are minimized as follows:
\begin{equation}\label{eq:inp}
\min_{c, \bm{g}, \bm{H}} = \Vert \bm{H}^{[h]} - \bm{H}^{[h-1]} \Vert_F^2 ~~ \text{subject~to} ~~ \bm{H} = \bm{H}^T ~~ \text{and} ~~ m[\bm{z}_k - \bm{\theta}(\tau)] = \Phi_{\gamma}[\bm{z}_k], \quad \forall k = 0, \ldots, \kappa,
\end{equation}
where $v+1 < \kappa + 1 < \frac{(v+1)(v+2)}{2}$ denotes the interpolation points, including the current iteration step, $\bm{H}^{[h-1]}$ is the Hessian of the previous model, which initialized as the zero matrices, and $\Vert \cdot \Vert_F^2$ is the Frobenius norm. The interpolation problem in~\eqref{eq:inp} provides a linear system of size $\kappa + v + 2$ so that the model is considered linear when $\kappa = v + 2$. As mentioned above, at each iteration, the BOBYQA method updates the interpolation methods by inserting a new point and removing the point that negatively affects model accuracy. The point to be removed from the model is determined based on the poisedness and stability concepts \citep[see, e.g.,][]{Powell2009, Cartis2022}.

The proposed method with the BOBYQA algorithm requires optimal values of the penalty parameters $\lambda_1$ and $\lambda_2$ to provide efficient estimation results. Several information criteria, such as BIC \citep{Koenker1994}, approximate cross-validation \citep{Nychka1995}, generalized cross-validation \citep{Yuan2006}, multifold cross-validation \citep{Zhang1993}, and the BIC \citep{Lee2014} can be used with the proposed method to determine optimum values of the smoothing parameters $\lambda_1$ and $\lambda_2$. We suggest using BIC to determine the optimum values of the smoothing parameters because of its simplicity and computing time. Following the definition of BIC in \cite{Schwarz}, we obtain the following BIC to determine the optimum values of the smoothing parameters:
\begin{equation*}
\text{BIC}(\lambda_0, \lambda_1) = \ln \bigg \Vert \sum_{i=1}^n\widetilde{\rho}_{\tau, \gamma} \left[ \Y_i(t) - \widehat{q}_{\lambda_1, \lambda_2} (\X_i) \right] \bigg \Vert_{\mathcal{L}_2} + \ln(n),
\end{equation*}
where $\widehat{q}_{\lambda_1, \lambda_2} (\X_i)$ denotes the $\tau^\textsuperscript{th}$ conditional quantile of the functional response estimated by the proposed method with smoothing parameters $(\lambda_1$, $\lambda_2)$ and $\Vert \cdot \Vert$ is the $\mathcal{L}_2$ norm.

We consider a trimmed version of the smooth objective function and BIC to make the parameter estimation process robust to atypical observations. Let $[ \cdot ]: \mathbb{R} \rightarrow \mathbb{Z}$ be a function rounding the elements to the nearest integer and let $\mathcal{Z} = \lbrace \mathcal{Z} \subset \lbrace 1, \ldots, n \rbrace, \vert \mathcal{Z} \vert = \lbrace [\iota n] \rbrace$ for a given $\iota \in [0,1]$. Then, we define the trimmed version of~\eqref{eq:rf} as follows:
\begin{equation*}
\Phi_{\gamma}[\bm{\theta}(\tau), \mathcal{Z}] = \sum_{i \in \mathcal{Z}} \widetilde{\rho}_{\tau, \gamma} \left[ \Y_i(t) - \bm{\Pi}_i \bm{\theta}(\tau) \right] + \bm{P}(\lambda_1, \lambda_2) \bm{\theta}(\tau).
\end{equation*}
Similarly, using the trimmed set, we consider the trimmed version of the BIC as follows:
\begin{equation*}
\text{BIC}(\lambda_0, \lambda_1, \mathcal{Z}) = \ln \bigg \Vert \sum_{i \in \mathcal{Z}} \widetilde{\rho}_{\tau, \gamma} \left[ \Y_i(t) - \widehat{q}_{\lambda_1, \lambda_2} (\X_i) \right] \bigg \Vert_{\mathcal{L}_2} + \ln([\iota n]).
\end{equation*}
In our numerical calculations, we take $\iota = 0.8$ \citep[i.e., 20\% trimming proportion, which is commonly used in robust statistics, see, e.g.,][]{Wilcox}.

The optimal values of the penalty parameters $\lambda_1$ and $\lambda_2$ are determined via a two-dimensional standard grid-search approach with a given set of candidate values for $\lambda_1$ and $\lambda_2$. We note that besides~$\lambda_1$ and~$\lambda_2$, the performance of the proposed method is also affected by the tuning parameter~$\gamma$. Including an extra parameter into the grid-search approach, that is, using three-dimensional grid-search ($\lambda_1$, $\lambda_2$, and $\gamma$), may need to be more computationally efficient. Thus, we use a pre-determined~$\gamma$ value in our numerical analyses. 

As shown in \cite{Zheng2011}, small values of~$\gamma$ ensure that the smoothed check loss function and the original check function are close \citep[see, e.g., Lemma 2 of][]{Zheng2011}. We perform a Monte Carlo experiment to show the effect of $\gamma$ on the performance of the proposed method. As discussed in Section~\ref{sec:mc}, the finite-sample performance of the proposed method decreases as the value of $\gamma$ increases. Thus, we consider a fixed $\gamma = 0.005$ in our numerical analyses.

\section{Monte Carlo experiments}\label{sec:mc}

A series of Monte-Carlo experiments under different data generation processes are conducted to evaluate the estimation and predictive performance of the proposed method (``pflqr'', hereafter). The finite-sample performance of the proposed pflqr method is compared with those of the functional principal component regression (fpcr), functional partial least squares regression model of \cite{Zhou2021} (fplsr), functional linear quantile regression model of \cite{BSMMA} (flqr), functional partial quantile regression model of \cite{BSA2022}, and the penalized function-on-function linear regression model of \cite{ivanescu2015} (pffr). 

In the experiments, three data generation processes (DGP-I, DGP-II, and DGP-III) are considered to generate the datasets. In DGP-I, similar to \cite{ivanescu2015} and \cite{Cai2021}, a relatively smooth process is used to generate the data. In this case, the aim is to show the correctness of the proposed method. Compared with DGP-I, DGP-II uses a more wavy regression parameter function, resulting in a more fluctuated functional response. This DGP aims to show if penalization results are better for the fluctuated functional data than unpenalized quantile regression models (flqr and fpqr). In DGP-III, in contrast to the initial two data generation processes, we examine a scenario where error terms exhibit temporal dependence. The objective of incorporating DGP-III is to illustrate how the models under investigation in this study address situations where the assumption of independent error terms is not tenable. In addition, for the first two DGPs, the generated datasets are contaminated by outliers at 5\% and 10\% contamination levels to evaluate the robust nature of the proposed method. Thus, the quantile level is set throughout the experiments to $\tau = 0.5$. 

In the experiments, the functional predictor $\X_i(s)$ is generated at 50 equally spaced points in the interval $[0,1]$, that is $s \in \lbrace r/50: r = 1, \ldots, 50 \rbrace$, while the functional response $\Y_i(t)$ is generated at 60 equally spaced points in the interval $[0,1]$, that is $t \in \lbrace j/60: j = 1, \ldots, 60 \rbrace$. For DGP-I, the following process is used to generate the functional predictor:
\begin{equation*}
\X_i(s) = \sum_{r=1}^{10} \frac{1}{r^2} \left\lbrace \xi_{i1,r} \sqrt{2} \sin(r \pi s) + \xi_{i2,r} \sqrt{2} \cos(r \pi s) \right\rbrace,
\end{equation*}
where $\xi_{i1,r}$ and $\xi_{i2,r}$ for $r = 1, \ldots, 10$ are independently generated from the standard normal distribution. Then, the functional response is generated as follows:
\begin{equation*}
\Y_i(t) = \alpha(t) + \int_0^1 \X_i(s) \beta(s,t) ds + \epsilon_i(t),
\end{equation*}
where $\alpha(t) = 2 e^{-(t-1)^2}$, $\beta(s,t) = 4 \cos(2 \pi t) \sin(\pi s)$, and $\epsilon_i$ is a Gaussian white noise stochastic process where $\epsilon_i(t_j)$ for $j = 1, \ldots, 60$ are generated from independent normal distributions with mean-zero and variance $0.01^2$, i.e., $\epsilon_i(t_j) \sim N(0, 0.01^2)$. To obtain the outlier contaminated data in this DGP, the randomly selected 5\% and 10\% of the generated data are replaced with the observations generated by the following regression parameter functions: $\alpha^{*,1}(t) = 4 e^{-t^2}$ and $\beta^{*,1}(s,t) = 6 \sin(4 \pi t) \sin(2 \pi s)$. 

In DGP-II, the datasets are generated similarly to DGP-I but using the regression coefficient functions $\alpha(t) = 2 \sin(4 \pi t)$ and $\beta(s,t) = 4 \cos(4 \pi t) \sin(4 \pi s)$. In DGP-II, the following regression parameter functions are used to generate the outlying observations: $\alpha^{*,2}(t) = 4 \cos(8 \pi t)$ and $\beta^{*,2}(s,t) = 6 \cos(8 \pi t) \sin(6 \pi s)$. 

\begin{figure}[!htbp]
  \centering
  \includegraphics[width=4.5cm]{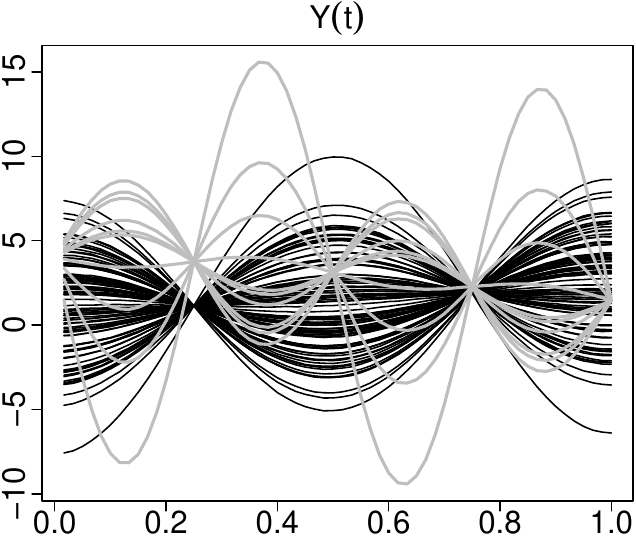}
  \includegraphics[width=4.5cm]{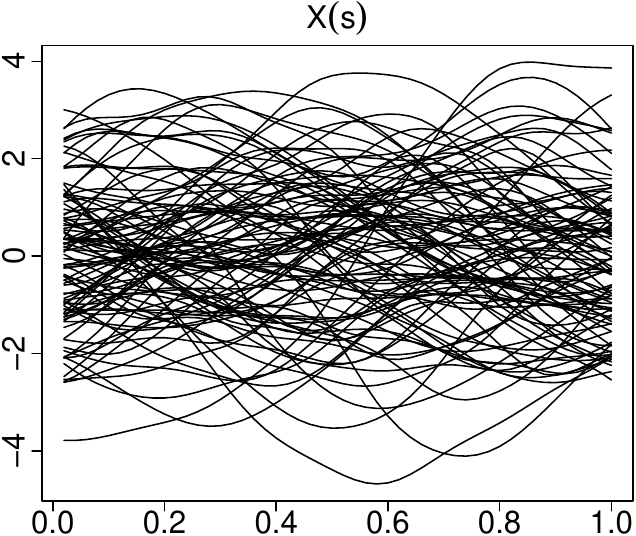}
  \includegraphics[width=4.5cm]{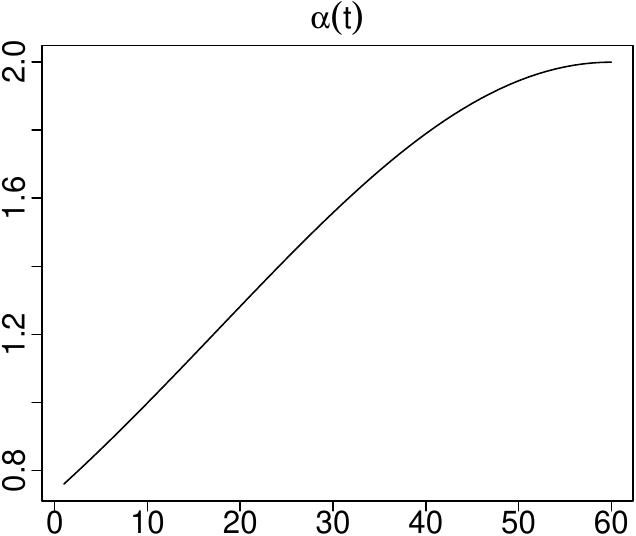}
  \includegraphics[width=3.4cm]{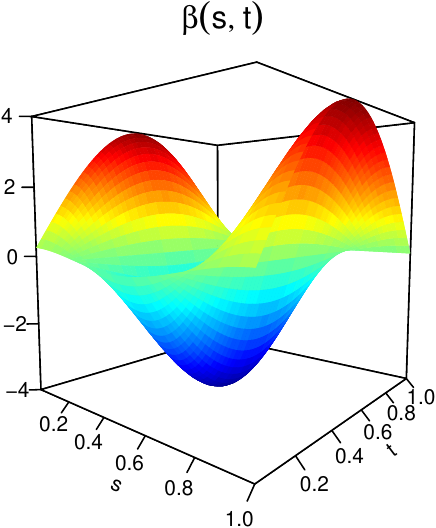}
  \\
  \includegraphics[width=4.5cm]{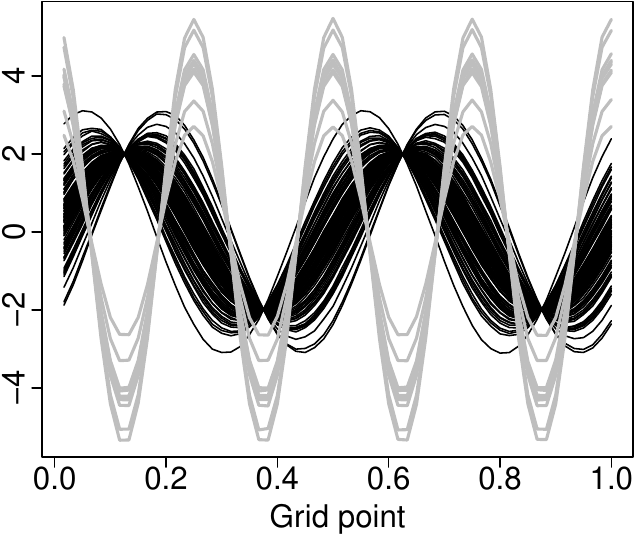}
  \includegraphics[width=4.5cm]{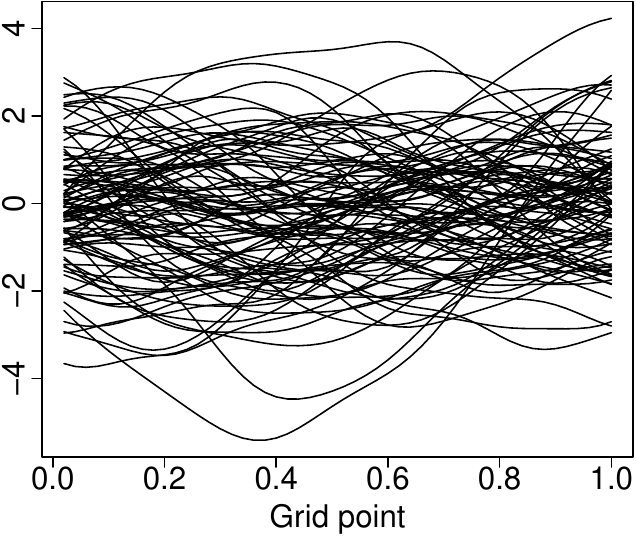}
  \includegraphics[width=4.5cm]{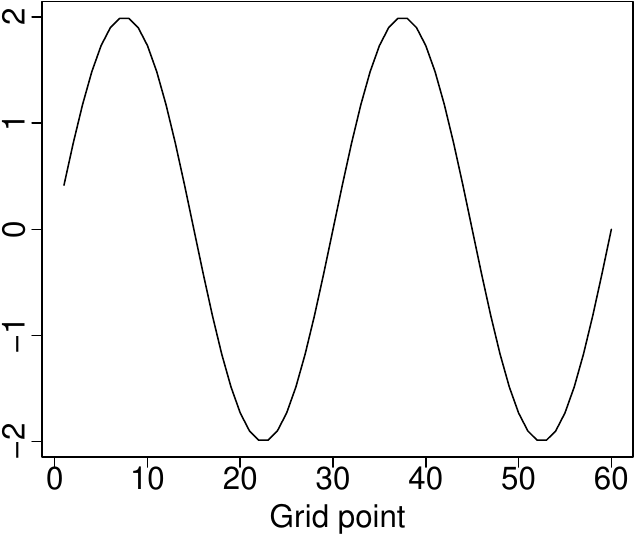}
  \includegraphics[width=3.4cm]{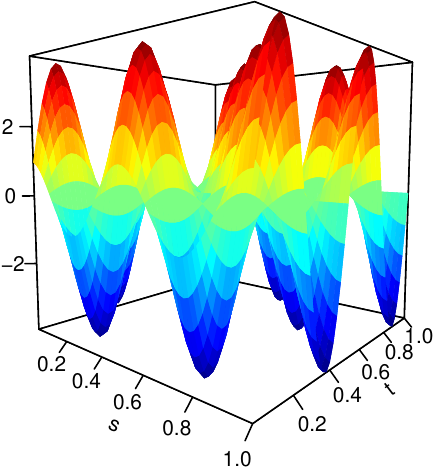}
  \caption{\small{Graphical display of the generated functional response (first column), functional predictor (second column), intercept function (third column), and the regression parameter function (fourth column) under DGP-I (first row) and DGP-II (second row). For the functional response (i.e., the first row), the observations generated by the DGP are shown as black curves, while the gray curves denote the outlying observations}.}\label{fig:Fig_1}
\end{figure}

In DGP-III, the datasets are generated similarly to DGP-II. However, in this particular DGP, the error process follows a multivariate normal distribution with a mean of zero and a variance-covariance matrix $\Sigma$, where $\Sigma$ is an $n \times n$ matrix with diagonal elements equal to 1 and off-diagonal elements set to $\rho = 0.8$. In DGP-III, the error terms exhibit a pronounced correlation structure. Figure~\ref{fig:Fig_1} displays a graphical representation of the generated data for both DGP-I and DGP-II. The data generated under DGP-III share a similar structure to that generated under DGP-II.

Throughout the experiments, for each DGP, three different sample sizes $n_{\text{train}} = [50, 100, 250]$ are considered. With the generated training data, the models are constructed, and the following root relative integrated squared percentage estimation errors (RISPEE) are computed for the intercept and regression parameter functions to evaluate the estimation performance of the methods:
\begin{align*}
\text{RRISPEE}(\widehat{\alpha}) &= 100 \times \sqrt{\frac{\Vert \alpha(t) - \widehat{\alpha}(t) \Vert_2^2}{\Vert \alpha(t) \Vert_2^2}}, \\
\text{RRISPEE}(\widehat{\beta}) &= 100 \times \sqrt{\frac{\Vert \beta(s,t) - \widehat{\beta}(s,t) \Vert_2^2}{\Vert \beta(s,t) \Vert_2^2}},
\end{align*}
where $\widehat{\alpha}(t)$ and $\widehat{\beta}(s,t)$ respectively denote the estimates of $\alpha(t)$ and $\beta(s,t)$ and $\Vert \cdot \Vert_2$ is the $\mathcal{L}_2$ norm. Note that the fpcr and fplsr methods assume that both the functional predictor and response variables are mean-zero processes and, thus, only the $\text{RRISPEE}(\widehat{\beta})$ is computed for these two methods. For each sample size, $n_{\text{test}} = 100$ independent samples are generated as the test sample. With the generated test samples, the predictive performance of the methods is evaluated by applying the fitted models based on the training sets to the test samples, and the following root mean squared percentage error (RMSPE) is computed:
\begin{equation*}
\text{RMSPE} = 100 \times \sqrt{\frac{\Vert \Y(t) - \widehat{\Y}(t) \Vert_2^2}{\Vert \Y(t) \Vert_2^2}},
\end{equation*}
where $\Y(t)$ is the true functional predictor in the test sample and $\widehat{\Y}(t)$ is its predicted value obtained by the methods. In addition, for the flqr, fpqr, and the proposed pflqr models, prediction intervals are constructed for the functional response in the test sample. First, the same model for two different quantile levels $\tau_1 = 0.025$ and $\tau_2 = 0.975$ are fitted to the training data. Then, the fitted models are applied to the test sample to construct a 95\% prediction interval for the response variable. 

The coverage probability deviance (CPD) and interval score (score) metrics are considered to evaluate the accuracy of the constructed prediction intervals. Let $Q_{\tau_1}$ and $Q_{\tau_2}$ denote the computed $2.5\%$ and $97.5\%$ quantiles of the functional response, respectively. Then, the CPD and score are defined as follows:
\begin{align*}
\text{CPD} &= \left| 0.95 - \frac{1}{100\times J}\sum^{100}_{i=1}\mathds{1} \sum_{j=1}^J \{Q_{\tau_{1}}\leq \Y_i(t_j)\leq Q_{\tau_{2}}\} \right| \\
\text{score} &= \frac{1}{100\times J} \sum_{i=1}^{100} \left| \lbrace Q_{\tau_2} - Q_{\tau_1} \rbrace + \frac{2}{0.05} \sum_{j=1}^J \lbrace Q_{\tau_1} - \Y_i(t_j) \rbrace \mathbb{1} \lbrace \Y_i(t_j) < Q_{\tau_1} \rbrace \right. \\
& \left. \hspace{2in}+ \frac{2}{0.05} \sum_{j=1}^J \lbrace \Y_i(t_j) - Q_{\tau_2} \rbrace \mathbb{1} \lbrace \Y_i(t_j) > Q_{\tau_2} \rbrace \right|,
\end{align*}
where $\{j = 1, \ldots, J \}$ denotes the time span where the response functions observed. The CPD is the absolute difference between the nominal and empirical coverage probabilities. Its small values correspond to the case where most of the observations in the test sample are covered by the prediction interval. On the other hand, the interval score is a metric used to evaluate coverage probability and width of the prediction interval simultaneously, and its small values correspond to accurate and narrower prediction intervals. 

We note that for each DGP and sample size, 100 Monte-Carlo replications are performed, and 15 tensor product $B$-spline basis expansion functions are used to construct the models for the pffr, flqr, and the proposed method, while the number of $B$-spline basis expansion functions used for the fpcr, fplsr, and fpqr are determined using cross-validation.

Before presenting the results, we first investigate the effect of the tuning parameter $\gamma$ on the finite-sample performance of the proposed method. For this purpose, we conduct a Monte-Carlo experiment under all the DGPs with 100 runs and compute the $\text{RRISPEE}(\widehat{\alpha})$, $\text{RRISPEE}(\widehat{\beta})$, and RMSPE values for different choices of $\gamma$ values ($\gamma = [0.005, 0.01, 0.05, 0.1, 0.25, 0.5]$). The results are presented in Table~\ref{tab:gamma}, which demonstrates that the finite-sample performance of the proposed method decreases as the value of $\gamma$ increases. Thus, we consider the fixed value $\gamma = 0.005$ in our numerical analyses.

\begin{small}
\begin{center}
\tabcolsep 0.23in
\begin{longtable}{@{}llcccccc@{}} 
\caption{\small{Computed mean $\text{RRISPEE}(\widehat{\alpha})$, $\text{RRISPEE}(\widehat{\beta})$, and RMSPE values over 100 Monte-Carlo replications for different $\gamma$ values}.}\label{tab:gamma} \\
\toprule
& & \multicolumn{6}{c}{$\gamma$} \\ \cmidrule{3-8}
{DGP} & Metric & 0.005 & 0.001 & 0.05 & 0.1 & 0.25 & 0.5 \\ \midrule
\endfirsthead
\toprule
{DGP} & Metric & \multicolumn{6}{c}{$\gamma$} \\ \midrule
& & 0.005 & 0.001 & 0.05 & 0.1 & 0.25 & 0.5 \\ \midrule
\endhead
\midrule
\multicolumn{8}{r}{Continued on next page} \\ 
\endfoot
\endlastfoot
DGP-I & $\text{RRISPEE}(\widehat{\alpha})$ & $0.610$ & $0.687$ & $0.897$ & $1.478$ & $3.488$ & $4.859$ \\
& $\text{RRISPEE}(\widehat{\beta})$ & $60.177$ & $62.266$ & $62.320$ & $63.276$ & $70.170$ & $76.635$  \\
& RMSPE & $5.053$ & $5.256$ & $5.275$ & $5.648$ & $7.645$ & $9.529$  \\
\cmidrule(l){2-8}

DGP-II & $\text{RRISPEE}(\widehat{\alpha})$ & $4.392$ & $8.388$ & $8.012$ & $8.115$ & $8.556$ & $8.974$ \\
& $\text{RRISPEE}(\widehat{\beta})$ & $67.949$ & $73.557$ & $72.009$ & $74.900$ & $73.413$ & $76.473$  \\
& RMSPE & $13.315$ & $14.861$ & $14.733$ & $15.106$ & $15.135$ & $15.748$  \\
\cmidrule(l){2-8}

DGP-III & $\text{RRISPEE}(\widehat{\alpha})$ & $14.992$ & $17.138$ & $17.751$ & $18.623$ & $20.348$ & $20.658$ \\
& $\text{RRISPEE}(\widehat{\beta})$ & $62.949$ & $64.563$ & $66.360$ & $66.668$ & $67.912$ & $68.153$  \\
& RMSPE & $15.311$ & $19.494$ & $19.625$ & $20.485$ & $21.642$ & $21.755$  \\
\bottomrule
\end{longtable}
\end{center}
\end{small}

The computed mean $\text{RRISPEE}(\widehat{\alpha})$, $\text{RRISPEE}(\widehat{\beta})$, RMSPE, CPD, and score values with their standard errors under DGP-I--DGP-III are presented in Tables~\ref{tab:tab_1}--~\ref{tab:tab_3}, respectively. From Table~\ref{tab:tab_1} (i.e., when a relatively smooth process is used to generate the data), the proposed method produces superior performance metrics over all of its competitors for all the performance metrics when no outlier is present in the data. While the unpenalized quantile regression models (flqr and fpqr) provide no superior performance than their unpenalized (fpcr and fplsr) and penalized (pffr) competitors, the proposed method produces superior performance over all of its competitors. This result shows that the derivative-free BOBYQA method improves the approximation of the regression parameters. Our results show that the pffr generally produces better results than the flqr and fpqr. This is because the penalization strategy of the pffr is used to control the smoothness of the estimates, which results in better estimates over the flqr and fpqr. 

When outliers are present in the data, the proposed method produces improved performance over all of its competitors except for $\text{RRISPEE}(\widehat{\beta})$. In this case, the flqr and fpqr produce better $\text{RRISPEE}(\widehat{\beta})$ values than the other methods. This result shows that the non-linear programming approach used in the flqr and fpqr to estimate the regression parameter function is more robust to outliers than the BOBYQA procedure used in the proposed method. The worst results for $\text{RRISPEE}(\widehat{\alpha})$ are generally produced by the flqr and fpqr because they use centered functional variables in their estimation steps, and the usual mean used for this purpose is not robust to outliers. In the case of outliers, the conditional mean regression models (i.e., pffr, fpcr, and fplsr) are generally more affected by outlying observations, and their performance worsens as the contamination level increases. On the other hand, compared with conditional mean regression models, the quantile regression models (i.e., flqr, fpqr, and pflqr, especially the proposed method) are less affected by outliers, and the decrease in their performance is bounded. Compared with flqr and fpqr, the proposed method produces improved prediction intervals regardless of sample size and contamination level; that is, the proposed method produces more accurate prediction intervals with narrower interval lengths for the functional response than those of flqr and fpqr.

\begin{small}
\begin{center}
\tabcolsep 0.2in
\begin{longtable}{@{}ccccccccc@{}} 
\caption{\small{Computed mean $\text{RRISPEE}(\widehat{\alpha})$, $\text{RRISPEE}(\widehat{\beta})$, RMSPE, CPD, and score values with their standard errors (given in brackets) over 100 Monte-Carlo replications under DGP-I. The results are obtained under three sample sizes ($n$) and contamination levels (\%) (0\% corresponds to outlier-free data generation case)}.}\label{tab:tab_1} \\
\toprule
{\%} & {n} & Method & $\text{RRISPEE}(\widehat{\alpha})$ & $\text{RRISPEE}(\widehat{\beta})$ & RMSPE & score & CPD \\ \midrule
\endfirsthead
\toprule
{\%} & {n} & Method & $\text{RRISPEE}(\widehat{\alpha})$ & $\text{RRISPEE}(\widehat{\beta})$ & RMSPE & score & CPD \\ \midrule
\endhead
\midrule
\multicolumn{8}{r}{Continued on next page} \\ 
\endfoot
\endlastfoot
$0\%$ & 50 & pffr & $0.153$ & $2.418$ & $0.750$ & -- & --  \\
& & & ($0.127$) & ($0.031$) & ($0.034$) & -- & --  \\

& & fpcr & -- & $4.483$ & $8.198$ & -- & --  \\
& & & -- & ($1.682$) & ($5.461$) & -- & -- \\

& & fplsr & -- & $1.511$ & $8.172$ & -- & --  \\
& & & -- & ($0.318$) & ($5.578$) & -- & -- \\

& & flqr & $11.938$ & $5.055$ & $12.268$ & 10.351 & $0.882$ \\
& &	& ($11.142$) & ($2.406$) & ($8.316$) & (7.601) & ($0.113$) \\

& & fpqr & $ 11.961$ & $2.613$ & $8.971$ & 7.242 & $0.900$ \\
& &	& ($11.138$) & ($1.438$) & ($8.196$) & (5.061) & ($0.087$) \\

& & pflqr & $0.068$ & $0.819$ & $0.229$ & $0.078$ & $0.042$\\
& & & ($0.022$) & ($0.089$) & ($0.032$) & ($0.011$) & ($0.005$) \\
\cmidrule(l){2-8}
					
& 100 & pffr & $0.132$ & $2.380$ & $0.748$ & -- & -- \\
& & & ($0.088$) & ($0.011$) & ($0.037$) & -- & -- \\

& & fpcr & -- & $2.546$ & $4.946$ & -- & --  \\
& & & -- & ($1.036$) & ($3.876$) & -- & -- \\

& & fplsr & -- & $1.073$ & $4.953$ & -- & --  \\
& & & -- & ($0.205$) & ($3.928$) & -- & -- \\
					
& & flqr & $10.416$ & $3.680$ & $9.924$ & 7.133 & $0.881$  \\
& &	& ($7.602$) & ($1.934$) & ($7.816$) & (6.414) & ($0.076$)  \\

& & fpqr & $10.417$ & $1.997$ & $7.746$ & 4.602 & $0.891$  \\
& &	& ($7.614$) & ($0.857$) & ($5.540$) & (3.578) & ($0.083$)  \\

& & pflqr & $0.051$ & $0.551$ & $0.168$ & $0.063$ & $0.044$\\
& & & ($0.023$) & ($0.061$) & ($0.013$) & ($0.003$) & ($0.002$) \\
\cmidrule(l){2-8}

& 250 & pffr & $0.083$ & $2.370$ & $0.747$ & -- & -- \\
& &	& ($0.060$) & ($0.005$) & ($0.033$) & -- & -- \\

& & fpcr & -- & $1.387$ & $2.499$ & -- & --  \\
& & & -- & ($0.619$) & ($1.965$) & -- & -- \\

& & fplsr & -- & $0.686$ & $2.485$ & -- & --  \\
& & & -- & ($0.122$) & ($1.998$) & -- & -- \\
					
& & flqr & $6.360$ & $2.949$ & $8.705$ & 6.510 & $0.883$ \\
& &	& ($5.175$) & ($1.279$) & ($6.994$) & (4.571) & ($0.092$) \\

& & fpqr & $6.353$ & $1.786$ & $4.789$ & 3.711 & $0.890$ \\
& &	& ($5.172$) & ($0.513$) & ($3.676$) & (2.432) & ($0.079$) \\
					
& & pflqr & $0.038$ & $0.363$ & $0.100$ & $0.058$ & $0.046$\\
& &	& ($0.023$) & ($0.046$) & ($0.0109$) & ($0.001$) & ($0.002$) \\
\cmidrule(l){1-8}

$5\%$ & 50 & pffr	& $7.367$ & $39.276$ & $9.237$ & -- & -- \\
& & & ($2.681$) & ($21.6650$) & ($3.415$) & -- & -- \\

& & fpcr & -- & $20.445$ & $12.881$ & -- & --  \\
& & & -- & ($22.405$) & ($5.556$) & -- & -- \\

& & fplsr & -- & $14.515$ & $12.468$ & -- & --  \\
& & & -- & ($19.120$) & ($5.604$) & -- & -- \\

& & flqr & $17.502$ & $2.146$ & $14.191$ & 36.880 & $0.497$ \\
& &	& ($11.219$) & ($0.835$) & ($9.340$) & (26.147) & ($0.176$) \\

& & fpqr & $15.265$ & $3.229$ & $11.279$ & 7.911 & $0.485$ \\
& &	& ($11.198$) & ($1.735$) & ($8.141$) & (3.584) & ($0.116$) \\

& & pflqr & $1.648$ & $26.070$ & $4.747$ & $1.273$ & $0.113$\\
& &	& ($0.971$) & ($15.129$) & ($2.140$) & ($1.221$) & ($0.048$) \\
\cmidrule(l){2-8}

& 100 & pffr & $7.683$ & $32.184$ & $8.440$ & -- & -- \\
& &	& ($1.561$) & ($17.015$) & ($1.869$) & -- & --  \\

& & fpcr & -- & $13.623$ & $10.489$ & -- & --  \\
& & & -- & ($11.766$) & ($3.238$) & -- & -- \\

& & fplsr & -- & $10.655$ & $10.637$ & -- & --  \\
& & & -- & ($8.402$) & ($3.005$) & -- & -- \\

& & flqr & $12.190$ & $1.325$ & $13.513$ & 37.561 & $0.566$ \\
& &	& ($7.618$) & ($0.554$) & ($8.386$) & (25.404) & ($0.240$) \\

& & fpqr & $10.210$ & $2.825$ & $7.608$ & 5.962 & $0.414$ \\
& &	& ($8.107$) & ($1.550$) & ($5.907$) & (2.192) & ($0.109$) \\

& & pflqr & $1.440$ & $23.262$ & $4.064$ & $0.826$ & $0.110$\\
& &	& ($0.756$) & ($8.054$) & ($1.330$) & ($0.343$) & ($0.033$) \\
\cmidrule(l){2-8}

& 250 & pffr	& $6.402$ & $22.346$ & $6.626$ & -- & -- \\
& & & ($0.783$) & ($12.122$) & ($1.025$) & -- & -- \\

& & fpcr & -- & $9.918$ & $6.990$ & -- & --  \\
& & & -- & ($7.856$) & ($1.591$) & -- & -- \\

& & fplsr & -- & $8.800$ & $7.049$ & -- & --  \\
& & & -- & ($9.181$) & ($1.544$) & -- & -- \\

& & flqr & $8.776$ & $0.852$ & $9.491$ & 32.729 & $0.480$ \\
& &	& ($4.226$) & ($0.318$) & ($6.058$) & (23.512) & ($0.210$) \\

& & fpqr & $6.246$ & $2.211$ & $4.714$ & 4.916 & $0.405$ \\
& &	& ($4.671$) & ($1.000$) & ($3.282$) & (1.655) & ($0.102$) \\

& & pflqr & $0.773$ & $14.481$ & $2.489$ & $0.501$ & $0.090$\\
& &	& ($0.297$) & ($3.902$) & ($0.533$) & ($0.169$) & ($0.024$) \\
\cmidrule(l){1-8}

$10\%$ & 50 & pffr & $15.406$ & $52.084$ & $16.804$ & -- & -- \\
& & & ($4.647$) & ($23.116$) & ($4.405$) & -- & -- \\

& & fpcr & -- & $24.707$ & $18.154$ & -- & --  \\
& & & -- & ($20.860$) & ($4.523$) & -- & -- \\

& & fplsr & -- & $28.183$ & $18.495$ & -- & --  \\
& & & -- & (63.832) & ($5.560$) & -- & -- \\

& & flqr & $20.886$ & $2.361$ & $17.476$ & 42.944 & $0.542$ \\
& &	& ($11.098$) & ($1.195$) & ($9.778$) & (29.423) & ($0.265$) \\

& & fpqr & $13.320$ & $4.902$ & $9.844$ & 8.924 & $0.399$ \\
& &	& ($11.464$) & ($3.849$) & ($8.281$) & (2.605) & ($0.103$) \\

& & pflqr & $3.599$ & $42.492$ & $8.229$ & $1.870$ & $0.118$\\
& &	& ($3.235$) & ($19.167$) & ($3.152$) & ($0.850$) & ($0.034$) \\
\cmidrule(l){2-8}

& 100 & pffr & $13.681$ & $40.868$ & $13.767$ & -- & -- \\
& &	& ($2.043$) & ($19.447$) & ($2.003$) & -- & -- \\

& & fpcr & -- & $19.119$ & $14.415$ & -- & --  \\
& & & -- & ($16.757$) & ($2.619$) & -- & -- \\

& & fplsr & -- & $15.207$ & $14.504$ & -- & --  \\
& & & -- & (9.083) & ($2.605$) & -- & -- \\
					
& & flqr & $15.695$ & $1.545$ & $13.059$ & 41.678 & $0.469$ \\
& & & ($6.835$) & ($0.620$) & ($6.301$) & (21.709) & ($0.205$) \\

& & fpqr & $8.521$ & $3.271$ & $6.331$ & 7.359 & $0.332$ \\
& &	& ($6.731$) & ($2.037$) & ($4.774$) & (2.081) & ($0.070$) \\
					
& & pflqr & $1.968$ & $30.964$ & $5.507$ & $1.233$ & $0.103$\\
& &	& ($0.997$) & ($10.027$) & ($1.397$) & ($0.427$) & ($0.024$) \\
\cmidrule(l){2-8}

& 250 & pffr & $12.536$ & $33.949$ & $12.741$ & -- & -- \\
& &	& ($0.971$) & ($14.449$) & ($1.195$) & -- & -- \\

& & fpcr & -- & $16.015$ & $13.143$ & -- & --  \\
& & & -- & ($9.025$) & ($1.436$) & -- & -- \\

& & fplsr & -- & $14.461$ & $13.122$ & -- & --  \\
& & & -- & ($7.716$) & ($1.426$) & -- & -- \\

& & flqr & $13.306$ & $1.065$ & $13.337$ & 39.266 & $0.447$ \\
& &	& ($5.736$) & ($0.502$) & ($6.073$) & (20.546) & ($0.164$)  \\

& & fpqr & $6.544$ & $2.779$ & $4.962$ & 6.395 & $0.299$ \\
& &	& ($5.181$) & ($1.958$) & ($3.741$) & (1.136) & ($0.048$) \\
 
& & pflqr & $1.398$ & $20.347$ & $4.076$ & $0.738$ & $0.083$\\
& &	& ($0.719$) & ($7.081$) & ($0.880$) & ($0.011$) & ($0.016$) \\
\bottomrule
\end{longtable}
\end{center}
\end{small}

From Table~\ref{tab:tab_2}, the proposed method yields superior results in terms of $\text{RRISPEE}(\widehat{\beta})$ and RMSPE compared to the other methods, irrespective of the presence of outliers and for all sample sizes. Conversely, the proposed method yields either similar or inferior $\text{RRISPEE}(\widehat{\alpha})$ values compared to the other methods. Notably, the pffr method produces the best results for $\text{RRISPEE}(\widehat{\alpha})$ in this context. Compared with DGP-I, DGP-II results in more complex functional data; that is, DGP-II uses a more wavy regression parameter function, which results in a more fluctuating functional response. 

In comparison with the results presented in Table~\ref{tab:tab_1}, the results presented in Table~\ref{tab:tab_2} indicate that the proposed method offers superior estimations for $\text{RRISPEE}(\widehat{\beta})$ and RMSPE in function-on-function regression, outperforming its competitors particularly in the case of complex (fluctuating) datasets. Among other methods, the functional partial least squares-based quantile regression model (fpqr) provides estimations for $\text{RRISPEE}(\widehat{\beta})$ and RMSPE that are comparable to those obtained by the proposed method. When outliers are present in the generated data, both the fpqr and proposed methods yield similar $\text{RRISPEE}(\widehat{\beta})$ values. However, the proposed method outperforms fpqr across all contamination levels and sample sizes. Additionally, as shown in Table~\ref{tab:tab_2}, both the proposed method and fpqr yield significantly improved CPD and score values compared to the flqr method. In contrast to fpqr, the proposed method yields smaller interval score values but larger CPD values. This indicates that the proposed method produces narrower prediction intervals than fpqr, resulting in improved CPD values for the fpqr method.

\begin{small}
\begin{center}
\tabcolsep 0.22in
\renewcommand{\arraystretch}{0.979}
\begin{longtable}{@{}cccccccc@{}} 
\caption{\small{Computed mean $\text{RRISPEE}(\widehat{\alpha})$, $\text{RRISPEE}(\widehat{\beta})$, RMSPE, CPD, and score values with their standard errors (given in brackets) over 100 Monte-Carlo replications under DGP-II. The results are obtained under three sample sizes ($n$) and contamination levels (\%) (0\% corresponds to outlier-free data generation case)}.}\label{tab:tab_2} \\
\toprule
{\%} & {n} & Method & $\text{RRISPEE}(\widehat{\alpha})$ & $\text{RRISPEE}(\widehat{\beta})$ & RMSPE & score & CPD \\ \midrule
\endfirsthead
\toprule
{\%} & {n} & Method & $\text{RRISPEE}(\widehat{\alpha})$ & $\text{RRISPEE}(\widehat{\beta})$ & RMSPE & score & CPD \\ \midrule
\endhead
\midrule
\multicolumn{8}{r}{Continued on next page} \\ 
\endfoot
\endlastfoot
$0\%$ & 50 & pffr & $5.145$ & $312.027$ & $21.960$ & -- & -- \\
& & & ($1.154$) & ($43.642$) & ($1.174$) & -- & --  \\

& & fpcr & -- & 93.929 & $18.389$ & -- & --  \\
& & & -- & ($0.513$) & ($1.339$) & -- & -- \\

& & fplsr & -- & 94.301 & $22.146$ & -- & --  \\
& & & -- & ($1.296$) & ($1.782$) & -- & -- \\

& & flqr & $7.565$ & $95.301$ & $18.343$ & 17.562 & $0.946$ \\
& &	& ($2.995$) & ($1.569$) & ($4.249$) & (2.103) & ($0.003$)  \\

& & fpqr & $7.844$ & 72.232 & $50.170$ & 1.951 & $0.074$ \\
& &	& ($2.891$) & ($9.268$) & ($3.331$) & ($0.226$) & ($0.012$) \\

& & pflqr & $5.216$ & $70.277$ & $14.521$ & $1.000$ & $0.155$\\
& &	& ($3.010$) & ($7.018$) & ($2.079$) & ($0.115$) & ($0.031$) \\
\cmidrule(l){2-8}

& 100 & pffr & $3.691$ & $225.796$ & $20.511$ & -- & -- \\
& &	& ($0.936$) & ($54.421$) & ($1.064$) & -- & -- 	\\

& & fpcr & -- & 93.909 & $16.164$ & -- & --  \\
& & & -- & ($0.298$) & ($1.080$) & -- & -- \\

& & fplsr & -- & 93.484 & $18.643$ & -- & --  \\
& & & -- & ($1.123$) & ($1.057$) & -- & -- \\
				
& & flqr & $5.457$ & $94.919$ & $15.790$ & 17.566 & $0.946$ \\
& &	& ($2.287$) & ($0.689$) & ($1.417$) & (1.631) & ($0.002$) \\

& & fpqr & $6.283$ & 65.872 & $49.967$ & 1.753 & $0.052$ \\
& &	& ($2.667$) & ($5.660$) & ($3.091$) & ($0.091$) & ($0.004$) \\
				
& & pflqr & $5.165$ & $60.125$ & $13.689$ & $0.679$ & $0.161$\\
& &	& ($6.145$) & ($5.194$) & ($3.867$) & ($0.109$) & ($0.031$) \\
\cmidrule(l){2-8}

& 250 & pffr 	& $2.536$ & $190.094$ & $19.412$ & -- & -- \\
& &	& ($0.727$) & ($12.044$) & ($0.824$) & -- & -- \\

& & fpcr & -- & 93.411 & $14.610$ & -- & --  \\
& & & -- & ($2.387$) & ($0.886$) & -- & -- \\

& & fplsr & -- & 92.413 & $15.706$ & -- & --  \\
& & & -- & ($1.549$) & ($0.873$) & -- & -- \\
				
& & flqr & $3.589$ & $94.571$ & $14.779$ & 16.237 & $0.946$ \\
& &	& ($1.520$) & ($0.405$) & ($1.205$) & (4.120) & ($0.005$) \\	

& & fpqr & $4.446$ & 66.106 & $48.572$ & 1.695 & $0.050$ \\
& &	& ($1.361$) & ($3.369$) & ($3.043$) & ($0.039$) & ($0.001$) \\
			
& & pflqr & $5.301$ & $60.153$ & $13.903$ & $0.430$ & $0.161$\\
& & & ($9.138$) & ($9.027$) & ($6.441$) & ($0.060$) & ($0.025$) \\
\cmidrule(l){1-8}

$5\%$ & 50 & pffr	& $10.343$ & $329.529$ & $24.043$ & -- & -- \\
& & & ($1.943$) & ($39.972$) & ($1.282$) & -- & -- \\

& & fpcr & -- & 94.972 & $22.061$ & -- & --  \\
& & & -- & ($5.415$) & ($3.737$) & -- & -- \\

& & fplsr & -- & 94.760 & $25.687$ & -- & --  \\
& & & -- & ($1.364$) & ($2.785$) & -- & -- \\

& & flqr & $8.371$ & $93.258$ & $18.857$ & 60.440 & $0.974$ \\
& & & ($3.508$) & ($5.849$) & ($4.187$) & (13.296) & ($0.032$) \\

& & fpqr & $8.883$ & 73.432 & $50.611$ & 3.384 & $0.087$ \\
& &	& ($3.134$) & ($7.310$) & ($3.225$) & ($0.612$) & ($0.019$) \\

& & pflqr & $10.4696$ & $72.857$ & $17.873$ & $1.572$ & $0.127$\\
& & & ($2.917$) & ($6.237$) & ($2.652$) & ($0.392$) & ($0.029$) \\
\cmidrule(l){2-8}

& 100 	& pffr	& $11.799$ & $298.602$ & $24.109$ & -- & -- \\
& & & ($1.426$) & ($69.871$) & ($1.109$) & -- & -- \\

& & fpcr & -- & 94.406 & $20.778$ & -- & --  \\
& & & -- & ($2.917$) & ($1.700$) & -- & -- \\

& & fplsr & -- & 93.618 & $22.762$ & -- & --  \\
& & & -- & ($1.038$) & ($1.279$) & -- & -- \\
					
& & flqr & $5.603$ & $94.080$ & $16.178$ & 67.153 & $0.964$ \\
& & & ($2.126$) & ($1.346$) & ($1.321$) & (15.048) & ($0.401$) \\

& & fpqr & $6.209$ & 69.379 & $48.419$ & 3.199 & $0.049$ \\
& &	& ($2.510$) & ($6.169$) & ($3.487$) & ($0.322$) & ($0.007$) \\
					
& & pflqr & $8.996$ & $63.601$ & $16.392$ & $1.080$ & $0.099$\\
& & & ($4.666$) & ($7.472$) & ($4.085$) & ($0.188$) & ($0.020$) \\
\cmidrule(l){2-8}

& 250 	& pffr	& $10.718$ & $208.047$ & $22.374$ & -- & -- \\
& & & ($0.728$) & ($27.777$) & ($0.927$) & -- & -- \\		

& & fpcr & -- & 93.888 & $18.585$ & -- & --  \\
& & & -- & ($0.186$) & ($1.010$) & -- & -- \\

& & fplsr & -- & 93.149 & $19.423$ & -- & --  \\
& & & -- & ($0.947$) & ($1.066$) & -- & -- \\			
					
& & flqr & $3.681$ & $94.364$ & $15.126$ & 70.287 & $0.976$ \\
& & & ($1.200$) & ($0.720$) & ($1.436$) & (12.957) & ($0.043$) \\

& & fpqr & $4.612$ & 65.505 & $48.682$ & 3.109 & $0.041$ \\
& &	& ($1.752$) & ($3.272$) & ($2.949$) & ($0.264$) & ($0.003$) \\
					
& & pflqr & $9.428$ & $62.439$ & $15.014$ & $0.799$ & $0.089$\\
& & & ($6.358$) & ($7.161$) & ($4.880$) & ($0.150$) & ($0.017$) \\
\cmidrule(l){1-8}

$10\%$ & 50 & pffr	& $21.815$ & $361.664$ & $30.514$ & -- & -- \\
& & & ($2.416$) & ($43.928$) & ($1.439$) & -- & -- \\

& & fpcr & -- & 95.513 & $29.871$ & -- & --  \\
& & & -- & ($6.487$) & ($3.345$) & -- & -- \\

& & fplsr & -- & 95.172 & $32.245$ & -- & --  \\
& & & -- & ($1.281$) & ($2.840$) & -- & -- \\

& & flqr & $8.902$ & $93.046$ & $20.075$ & 62.630 & $0.954$ \\
& & & ($3.872$) & ($6.116$) & ($6.175$) & (23.665) & ($0.020$) \\

& & fpqr & $9.292$ & 75.241 & $50.677$ & 4.128 & $0.085$ \\
& &	& ($2.944$) & ($9.795$) & ($3.605$) & ($0.610$) & ($0.020$) \\

& & pflqr	& $15.605$ & $74.103$ & $21.461$ & $1.790$ & $0.103$\\
& & & ($2.758$) & ($7.123$) & ($3.089$) & ($0.445$) & ($0.023$) \\
\cmidrule(l){2-8}

& 100 & pffr	& $21.957$ & $357.713$ & $30.319$ & -- & -- \\
& & & ($1.994$) & ($50.599$) & ($1.548$) & -- & -- \\

& & fpcr & -- & 94.898 & $27.741$ & -- & --  \\
& & & -- & ($3.507$) & ($2.433$) & -- & -- \\

& & fplsr & -- & 94.094 & $29.217$ & -- & --  \\
& & & -- & ($0.782$) & ($2.230$) & -- & -- \\
					
& & flqr & $6.195$ & $94.053$ & $16.552$ & 65.272 & $0.952$ \\
& & & ($2.805$) & ($1.406$) & ($2.134$) & (24.372) & ($0.015$) \\

& & fpqr & $7.098$ & 71.399 & $49.375$ & 3.859 & $0.05$ \\
& &	& ($1.883$) & ($6.146$) & ($3.895$) & ($0.376$) & ($0.004$) \\
					
& & pflqr	& $11.7924$ & $67.952$ & $17.741$ & $1.214$ & $0.087$\\
& & & ($5.040$) & ($7.772$) & ($4.872$) & ($0.246$) & ($0.019$) \\
\cmidrule(l){2-8}

& 250 	& pffr	& $21.950$ & $233.040$ & $29.152$ & -- & -- \\
& & & ($1.067$) & ($43.836$) & ($1.021$) & -- & -- \\

& & fpcr & -- & 94.376 & $26.282$ & -- & --  \\
& & & -- & ($2.727$) & ($1.151$) & -- & -- \\

& & fplsr & -- & 93.440 & $26.926$ & -- & --  \\
& & & -- & ($0.624$) & ($1.066$) & -- & -- \\
					
& & flqr & $4.113$ & $94.437$ & $15.194$ & 65.155 & $0.951$ \\
& & & ($1.427$) & ($0.549$) & ($1.327$) & (25.165) & ($0.012$) \\

& & fpqr & $5.609$ & 66.230 & $48.100$ & 3.869 & $0.038$ \\
& &	& ($1.227$) & ($3.941$) & ($3.160$) & ($0.233$) & ($0.003$) \\
					
& & pflqr & $16.228$ & $65.768$ & $16.741$ & $0.889$ & $0.079$\\
& & & ($3.717$) & ($6.951$) & ($3.201$) & ($0.146$) & ($0.016$) \\
\bottomrule
\end{longtable}
\end{center}
\end{small}

In comparison to the results in Table~\ref{tab:tab_2} under no contamination, Table~\ref{tab:tab_3} shows that all methods, including the proposed approach, generally exhibit inferior performance when the errors possess a strong correlation structure. Among the other models, the proposed method demonstrates enhanced performance in estimation and prediction, i.e., it produces less $\text{RRISPEE}(\widehat{\alpha})$, $\text{RRISPEE}(\widehat{\beta})$, and RMSPE values than its competitors for all the sample sizes. 

Among the quantile regression models, the largest CPD and score values are produced by the flqr. Compared with the proposed method, fpqr yields significantly improved CPD and score values. This discrepancy arises because the proposed approach tends to generate narrower prediction intervals for the functional response. Consequently, most of the functional response values in the test sample are not encompassed by the prediction intervals constructed by the proposed approach, resulting in elevated CPD and interval score values.

\begin{small}
\begin{center}
\centering
\tabcolsep 0.26in
\renewcommand{\arraystretch}{0.934}
\begin{longtable}{@{}cccccccc@{}} 
\caption{\small{Computed mean $\text{RRISPEE}(\widehat{\alpha})$, $\text{RRISPEE}(\widehat{\beta})$, RMSPE, CPD, and score values with their standard errors (given in brackets) over 100 Monte-Carlo replications under DGP-III. The results are obtained under three sample sizes ($n$) when the errors follow a multivariate normal distribution}.}\label{tab:tab_3} \\
\toprule
$n$ & Method & $\text{RRISPEE}(\widehat{\alpha})$ & $\text{RRISPEE}(\widehat{\beta})$ & RMSPE & score & CPD \\ \midrule
\endfirsthead
\toprule
$n$ & Method & $\text{RRISPEE}(\widehat{\alpha})$ & $\text{RRISPEE}(\widehat{\beta})$ & RMSPE & score & CPD \\ \midrule
\endhead
\midrule
\multicolumn{8}{r}{Continued on next page} \\ 
\endfoot
\endlastfoot
50 & pffr & $29.682$ & $322.697$ & $35.612$ & -- & -- \\
& & ($6.637$) & ($31.907$) & ($4.912$) & -- & --  \\

& fpcr & -- & $93.993$ & $60.187$ & -- & --  \\
& & -- & ($0.559$) & ($4.929$) & -- & -- \\

& fplsr & -- & $94.299$ & $61.646$ & -- & --  \\
& & -- & ($0.988$) & ($4.763$) & -- & -- \\

& flqr & $31.474$ & $95.099$ & $34.576$ & 22.249 & $0.830$ \\
& & ($6.483$) & ($1.658$) & ($5.693$) & (3.807) & ($0.080$)  \\

& fpqr & $84.247$ & 88.877 & $80.490$ & 3.269 & $0.130$ \\
& & ($1.741$) & ($2.061$) & ($1.725$) & ($0.826$) & ($0.054$) \\

& pflqr & $27.189$ & 64.879 & $29.020$ & 13.266 & $0.749$\\
& & ($6.981$) & ($6.906$) & ($5.873$) & ($2.722$) & ($0.005$) \\
\midrule

100 & pffr & $30.892$ & $313.523$ & $36.074$ & -- & -- \\
& & ($5.612$) & ($45.766$) & ($4.410$) & -- & --  \\

& fpcr & -- & $93.850$ & $60.179$ & -- & --  \\
& & -- & ($0.304$) & ($5.336$) & -- & -- \\

& fplsr & -- & $93.574$ & $60.830$ & -- & --  \\
& & -- & ($0.767$) & ($5.232$) & -- & -- \\

& flqr & $31.875$ & $94.744$ & $33.509$ & 21.530 & $0.788$ \\
& & ($5.409$) & ($0.699$) & ($4.582$) & (3.992) & ($0.101$)  \\

& fpqr & $83.945$ & 87.133 & $80.162$ & 2.866 & $0.090$ \\
& & ($1.246$) & ($0.803$) & ($1.174$) & ($0.419$) & ($0.025$) \\

& pflqr & $26.062$ & 60.474 & $26.964$ & 11.553 & $0.813$\\
& & ($5.096$) & ($5.183$) & ($4.277$) & ($2.38$) & ($0.001$) \\
\midrule

250 & pffr & $29.788$ & $194.491$ & $34.595$ & -- & -- \\
& & ($5.924$) & ($14.903$) & ($4.524$) & -- & --  \\

& fpcr & -- & $93.387$ & $59.598$ & -- & --  \\
& & -- & ($2.604$) & ($4.496$) & -- & -- \\

& fplsr & -- & $92.803$ & $59.906$ & -- & --  \\
& & -- & ($1.217$) & ($4.453$) & -- & -- \\

& flqr & $30.141$ & $94.609$ & $31.985$ & 21.694 & $0.803$ \\
& & ($6.063$) & ($0.393$) & ($5.126$) & (3.915) & ($0.121$)  \\

& fpqr & $84.198$ & 86.252 & $80.088$ & 2.190 & $0.060$ \\
& & ($1.742$) & ($0.452$) & ($1.778$) & ($0.443$) & ($0.022$) \\

& pflqr & $24.477$ & 58.406 & $24.916$ & 12.770 & $0.950$\\
& & ($5.867$) & ($3.591$) & ($5.059$) & ($2.442$) & ($0.001$) \\
\bottomrule
\end{longtable}
\end{center}
\end{small}

\section{Empirical data analysis: Mary river flow data}\label{sec:data}

\begin{figure}[!htbp]
  \centering
  \includegraphics[width=8.9cm]{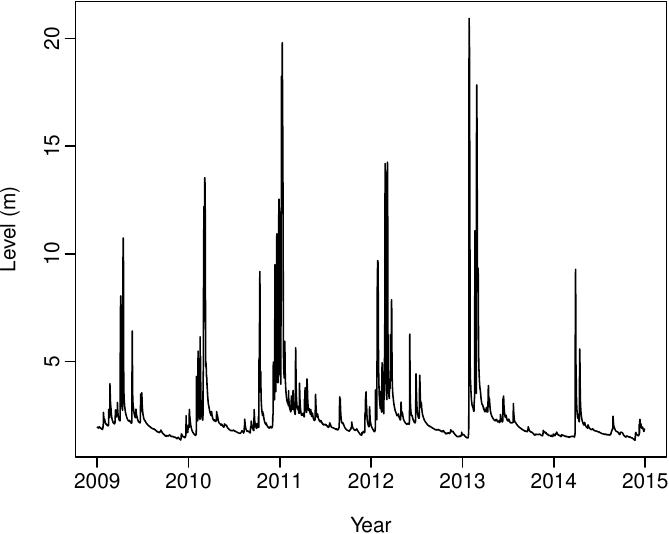}
  \quad
  \includegraphics[width=8.45cm]{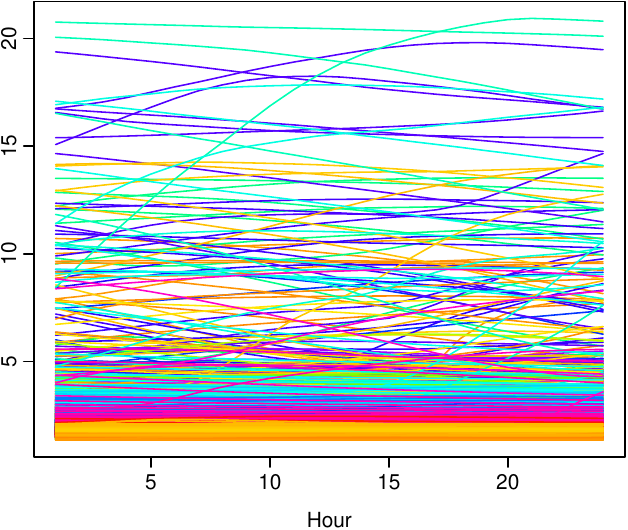}
  \caption{\small{Graphical display of the classical (left panel) and functional (right panel) time series of the river flow observations in the Mary River}.}\label{fig:Fig_2}
\end{figure}

We consider the Mary River flow dataset available in the \Rlogo \ package ``\texttt{robflreg}'' \citep{robflreg, BSR}. The data include hourly river flow observations for the Mary River, Australia, from January 2009 to December 2014 (six years). For the dataset, the observations are assumed to be a function of hours; that is, there are $n = 2190$ functional observations in total $\lbrace \Y_i(t): 1 \leq t \leq 24,~ i = 1, \ldots, 2190 \rbrace$. A graphical display of the traditional and functional time series of the river flow observations is presented in Figure~\ref{fig:Fig_2}.

With this dataset, we aim to evaluate how the previous river-flow curve time series $\Y_{i-1}(t)$ affects the current river-flow curve time series $\Y_i(t)$. For this purpose, we consider the autoregressive Hilbertian process of order one (ARH(1)), which is the functional analog of the well-known autoregressive process of one model. In more detail, we assume the following ARH(1) process:
\begin{equation*}
Y_i(t) = \alpha(t) + \int_1^{24} Y_{i-1}(s) \beta(s,t) ds dt, + \epsilon_i(t) \quad \forall t,s \in [1,24], \quad i = 1, \ldots, n-1,
\end{equation*}
where $\epsilon_i(t)$ is the error process. Let $\widehat{\alpha}(t)$ and $\widehat{\beta}(s,t)$ denote the estimates of $\alpha(t)$ and $\beta(s,t)$, respectively. Then, the forecast of the $(i+1)^\textsuperscript{th}$ river flow curve time series is obtained as follows: 
\begin{equation*}
\widehat{Y}_{i+1}(t) = \widehat{\alpha}(t) + \int_{s=1}^{24} Y_{i}(s) \widehat{\beta}(s,t)ds.
\end{equation*}
From Figure~\ref{fig:Fig_2}, it is clear that the Mary River flow data have heteroskedastic and outlying flow level observations. Thus, the proposed method produces improved results over the conditional mean regression models, that is pffr, fpcr, and fplsr. In addition, because the proposed method uses a penalization procedure, it is also expected to produce improved results over the unpenalized quantile regression models, that is flqr and fpqr.

We consider an expanding-window approach to evaluate the predictive performance of the methods. The data are divided into a training sample of days from 1/1/2009 to 09/11/2014 (2139 days) and a test sample from 10/11/2014 to 29/12/2014 (50 days). The models are constructed based on the entire training sample to forecast river flow observations on 11/11/2014. Then, the river flow observation on 12/11/2014 is forecasted by increasing the training sample by one day. This process is repeated until all the observations in the test sample are forecasted (i.e., the training sample covers the entire dataset). For pffr, flqr, and pflqr, ten tensor product $B$-spline basis expansion functions are used to construct the models, while the number of basis expansion functions for the fpcr, fplsr and fpqr are determined using cross-validation. The quantile level is set to $\tau = 0.5$. The RMSPE metric is computed for each forecasted curve to compare the predictive performance of the methods. In addition, a 95\% prediction interval is constructed for the flqr and pflqr methods by fitting two quantile regression models (using quantile levels $\tau_1 = 0.025$ and $\tau_2 = 0.975$) to the training data. To evaluate the prediction intervals, the CPD and score metrics are computed.

The computed RMSPEs, along with standard errors given in brackets, are $3.588~(4.866)$, $5.691~(3.581)$, $5.558~(3.746)$, $49.936~(21.047)$, $4.310~(4.478)$, and $3.099~(4.290)$ for the pffr, fpcr, fplsr, flqr, fpqr, and pflqr methods, respectively. From these results, the proposed method produces improved predictive performance, that is smaller RMSPE values, than all of its competitors.

The computed mean interval score and CPD and values, as well as their standard errors (given in brackets), respectively, are $20.003~(9.294)$ and $0.895~(0.101)$ for the flqr, $0.979~(0.457)$ and $0.062~(60.064)$ for the fpqr, and they are $1.123~(2.542)$ and $0.597~(0.252)$ for the proposed pflqr method. These results indicate that the fpqr and the proposed pflqr methods produce more accurate prediction intervals with narrower interval lengths than the flqr method. The improved prediction interval results of fpqr over the proposed method are due to the prediction intervals constructed by the proposed method being narrower and covering fewer observations of future river flow curves than those of fpqr.

Moreover, using the Mary river flow data, we compare the predictive performance of only functional quantile regression models, that is flqr, fpqr, and pflqr, at different quantile levels. While doing so, the same expanding-window approach described above is considered. The flqr, fpqr, and pflqr methods are performed at quantile levels $\tau = [0.2, 0.3, 0.4, 0.6, 0.7, 0.8]$ to predict the future values of the river flow observations, and the RMSPE metric is computed to compare the performance of the methods. The results are presented in Table~\ref{tab:dq}. From this table, the proposed pflqr method produces smaller RMSPE values than flqr and fpqr except the last two upper quantile levels. As in the results obtained when $\tau = 0.5$, the flqr produces the worst RMSPE values.

\begin{table}[!htb]
\caption{\small{Computed mean RMSPE values of flqr, fpqr, and pflqr methods at different quantile levels for Mary river flow data}.}
\centering
\tabcolsep 0.33in
\begin{tabular}{@{}lrrrrrr@{}} 
\toprule
 & \multicolumn{6}{c}{$\tau$} \\ \cmidrule{2-7}
Method & 0.2 & 0.3 & 0.4 & 0.6 & 0.7 & 0.8 \\ \midrule
flqr & $46.242$ & $48.576$ & $49.557$ & $50.379$ & $50.852$ & $51.859$ \\
fpqr & $9.587$ & $6.609$ & $5.135$ & $3.743$ & $3.027$ & $2.278$ \\
pflqr & $3.196$ & $3.187$ & $2.463$ & $3.136$ & $3.138$ & $3.211$ \\
\bottomrule
\end{tabular}\label{tab:dq} 
\end{table}

\section{Conclusion}\label{sec:conc}

Quantile regression is a general model for the entire conditional distribution of the response variable for a given set of predictors. While it has been well studied in scalar-on-function regression models, few studies examine quantile regression in function-on-function regression models. The existing function-on-function quantile regression models are based on the functional principal component and partial least squares decomposition of the functional variables. While these methods are practical, they may not provide improved results for complex structured functional data. This is because, in these methods, the smoothness of the functional parameter is induced by the basis dimension of the functional predictor, leading to significant under-smoothing if the functional parameter is considerably smoother than the higher-order partial least squares and principal component scores.

In this paper, we propose a novel function-on-function penalized quantile regression method to characterize the entire conditional distribution of a functional response for a given functional predictor. In the proposed method, tensor cubic $B$-splines expansion is used to represent the parameter functions, and the quadratic roughness penalties are applied to the expansion coefficients to control the smoothness of the estimates. In addition, the derivative-free optimization algorithm BOBYQA is used to estimate the regression parameter functions.

The estimation and predictive performance of the proposed method are evaluated via a series of Monte-Carlo experiments and an empirical data analysis, and the results are compared favorably with existing methods. In Monte-Carlo experiments, the datasets following the conditional mean regression model are generated under three DGPs.
\begin{itemize}
\item[1)] The dataset is generated using a relatively smooth process to show the accuracy of the proposed method. When the underlying data is best suited for the conditional mean regression (as in the first DGP), the quantile regression is expected to provide a similar but not superior performance to that of conditional mean regression. However, our results demonstrate that the proposed method produces superior performance over its competitors; compared with existing methods, it provides better parameter estimates and predictions. In other words, our results indicate that the proposed method's estimation algorithm greatly improves the regression parameters' approximation. 
\item[2)] Compared with the first DGP, a more wavy regression parameter function is used to generate a more fluctuating functional response. Our records demonstrate that the proposed method produces significantly improved results compared to the existing conditional mean and quantile regression models for this case. In particular, our results in this case show the penalization step used in the proposed method results in better estimation and prediction results for the function-on-function quantile regression model. In addition, the prediction intervals for the proposed method are constructed by fitting the same model using different quantile levels. 
\item[3)] In addition to DGP-II, we consider correlated error terms when generating the data. In this case, the proposed method also produces improved estimation and prediction performance over its competitors. However, when the errors are correlated, the prediction intervals constructed by the proposed method fail to cover the values of the functional response variable and produce larger interval scores and CPD values.
\end{itemize}

The methodology presented in this study can be extended in several research directions. For instance, in the current study, we consider only one functional predictor in the model. However, multiple functional predictors may better characterize the functional response's conditional distribution, and the proposed method can be extended function-on-multiple-function linear regression model. Further, our numerical results show that the linear programming approach used in the traditional quantile regression setup is more robust to outliers for smooth functional observations than the optimization algorithm used in the proposed method. Thus, a robust optimization algorithm may be used with the proposed method to make the current methodology more robust to outliers. 

Our estimation approach for the function-on-function linear quantile regression model employs a smooth check function as the objective function, with the smoothness of the functional regression coefficient achieved through a $B$-spline basis expansion. While our method for estimating conditional quantiles is conceptually straightforward, as demonstrated by \cite{Liu2020}, it may be less effective for handling complex and irregular functions. Furthermore, it provides quantiles at the population level and may not be the most efficient means of estimation. In such scenarios, the Bayesian frameworks proposed by \cite{yang2020} and \cite{Liu2020} can complement our approach, allowing for the estimation of not only population-level quantiles, but also quantiles specific to an individual subject's distribution.

The spatial quantile regression approach introduced by \cite{Zhang2022} can be integrated with our method to estimate the conditional spatial distribution of a functional response given functional predictors. Moreover, incorporating the Bayesian framework introduced by \cite{Liu2020} into our modeling strategy can enhance the estimation and predictive performance of our proposed method.

One challenge with the considered quantile estimation strategy is that there may be high variability in the estimate at extreme quantile levels, stemming from insufficient data in the tails of the distribution. In such cases, the generalized regression quantiles, which depend on principal component functions proposed by \cite{Guo2015}, can be extended to our approach to enhance its accuracy in estimating extreme quantile levels. Alternatively, when the response variable consists of random functions, other methods such as depth-based quantiles \cite{cuevas2007, Fraiman2012, Serfling2017, Chowdhury2019} and the functional boxplot approach proposed by \cite{Sun2011} can be considered. These techniques can also be adapted to our model to improve its predictive performance. Finally, we consider the approximate check loss function introduced by \cite{Zheng2011} to overcome the differentiability problem of the original check loss function. Alternatively, the smooth objective function introduced by \cite{FGH21} and \cite{HPT23}, which is asymptotically differentiable, can be implemented into the proposed method to obtain a much-improved inference for the function-on-function linear quantile regression model.

\section*{Acknowledgment}

We would like to thank an Associate Editor and two reviewers for their careful reading of our manuscript and for their valuable suggestions and comments, which have helped us produce an improved version of our article. The second author thanks financial support from an Australian Research Council Discovery Project DP230102250.

\section*{Statements and Declarations}

\subsection*{Funding}
No funding was received to assist with the preparation of this manuscript.

\subsection*{Conflict of interest}
The authors have no competing interests to declare that are relevant to the content of this article.

\subsection*{Code availability}
Example \Rlogo \ scripts for the Monte-Carlo experiments and the empirical data example are available at \url{https://github.com/UfukBeyaztas/pflqr}.

\newpage
\bibliographystyle{agsm}
\bibliography{penqr}

\end{document}